%% file: ms.tex
\documentclass[12pt,preprint]{aastex}
\usepackage{rotating}

\slugcomment{Accepted in ApJ}

\shorttitle{Observations of HD~100546}
\shortauthors{Ardila et al.}

\begin{document}

\title{HST/ACS Coronagraphic Observations of the Dust Surrounding HD~100546\footnote{For the astro-ph version, the images have been reduced in size and resolution. Please contact D. Ardila for a full resolution PDF.}}


\author{
D.R. Ardila\altaffilmark{1,2}, D.A. Golimowski\altaffilmark{2}, J.E. Krist\altaffilmark{3}, M. Clampin\altaffilmark{4}, H.C. Ford\altaffilmark{2}, \& G.D. Illingworth\altaffilmark{5}}

\altaffiltext{1}{Spitzer Science Center, California Institute of Technology, MS 220-6, Pasadena, CA 91125; ardila@ipac.caltech.edu}

\altaffiltext{2}{Department of Physics and Astronomy, Johns Hopkins
University, Baltimore, MD 21218.}

\altaffiltext{3}{NASA Jet Propulsion Laboratory, Pasadena, CA 91109.}

\altaffiltext{4}{NASA Goddard Space Flight Center, Code 681, Greenbelt, MD 20771.}

\altaffiltext{5}{UCO/Lick Observatory, University of California, Santa Cruz, CA 95064.}

\begin{abstract}

We present ACS/HST coronagraphic observations of HD~100546, a B9.5 star, 103 pc away from the sun, taken in the F435W, F606W, and F814W bands. Scattered light is detected up to 14'' from the star. The observations are consistent with the presence of an extended flattened nebula with the same inclination as the inner disk. The well-known ``spiral arms'' are clearly observed and they trail the rotating disk material. Weaker arms never before reported are also seen. The inter-arm space becomes brighter, but the structures become more neutral in color at longer wavelengths, which is not consistent with models that assume that they are due to the effects of a warped disk. Along the major disk axis, the colors of the scattered-light relative to the star are $\Delta (F435W-F606W)\approx0.0$--0.2 mags and $\Delta (F435W-F814W)\approx0.5$--1 mags. To explain these colors, we explore the role of asymmetric scattering, reddening, and large minimum sizes on ISM-like grains. We conclude each of these hypotheses by itself cannot explain the colors. The disk colors are similar to those derived for Kuiper Belt objects, suggesting that the same processes responsible for their colors may be at work here. We argue that we are observing only the geometrically thick, optically thin envelope of the disk, while the optically thick disk responsible for the far-IR emission is undetected. The observed spiral arms are then structures on this envelope. The colors indicate that the extended nebulosity is not a remnant of the infalling envelope but reprocessed disk material.

\end{abstract}

\keywords{stars: pre-main sequence -- circumstellar matter -- planetary systems: protoplanetary disks --  stars: imaging --  stars: individual \objectname{HD100546}}

\section{Introduction}

Resolved images of circumstellar disks in optical scattered light provide a unique way of establishing the spatial distribution of circumstellar material. They allow for the highest spatial resolution of all techniques available and are sensitive to light scattered by small particles regardless of their temperature. However, at optical wavelengths one has to contend with the brightness of the host star and the use of a coronagraphic mask becomes a necessity. Scattered-light observations complement thermal observations, and constrain models based on spectroscopic data only.

In this paper, we present coronagraphic optical images in which the circumstellar material around the B9.5V Herbig Ae/Be star HD~100546 is resolved in scattered light. HD~100546 is a single star, with d=$103^{+7}_{-6}$ pc and reddening of A$_V$=0.28 \citep{van98}, located at the edge of the Lynds dark cloud DC~296.2-7.9. Isochrone fitting using the models developed by \citet{pal93} suggests its age is $\gtrsim$10 Myr \citep{van98}. The age cannot be much larger than this, as the system presents evidence of strong accretion in the H$_\alpha$ lines \citep*{vie99}. Its fractional excess luminosity is  $L_{IR}/L_*=0.51$ ($L_{IR}$ is the excess infrared luminosity and $L_*$ is the stellar luminosity, see \citealt{mee01}).  

The circumstellar material was first resolved from the ground in the $J$ and $K_s$ bands, using the Adaptive Optics Near Infrared System (ADONIS) mounted on the European Southern Observatory's 3.6 m telescope at La Silla \citep*{pan00}. Those coronagraphic observations revealed nebulosity at angular distances ranging from 0.''4 to  2'' from the star. The authors concluded that the observations were consistent with the presence of a disk with an inclination $50^{\circ}\pm5^{\circ}$ from edge-on. \citet{aug01} performed Near Infrared Camera and Multi-Object Spectrometer (NICMOS) coronagraphic observations at 1.6 $\mu$m and detected scattered light from $\sim$0.''5 to 3.''8 from the star, as well as significant azimuthal asymmetries in the circumstellar disk. The observations had a spatial resolution of $\sim$0.''2 \citep{mob04}. The highest spatial resolution images to date ($\sim$0.''1 per resolution element) have been published by \citet{gra01}. These are broadband coronagraphic observations taken with the Space Telescope Imaging Spectrograph (STIS) that probe angular distances as small as  $\sim$0.''8 from the star (at selected azimuthal angles). The observations reveal a complex circumstellar structure which the authors interpret as an inclined disk surrounded by an envelope of material extending up to $\sim$10'' from the star. In addition, the disk presents bright and dark lanes, reminiscent of spiral structure. Thermal observations at $\sim$10 $\mu$m using nulling interferometry reveal that the disk has a radius of 30 AU at that wavelength \citep{liu03}. Those observations also suggest the presence of an inner hole.

The spectral energy distribution (SED) of this system has been intensively studied. The star is an example of a class of Herbig Ae/Be objects that show substantial silicate emission at 10 $\mu$m and for which there is strong emission from the mid-infrared (mid-IR) relative to the near-infrared (near-IR). These have been called group Ia sources by \citet{mee01}. AB Aurigae (4 Myrs, A0V, see \citealt{van97, dewarf03}) is another group Ia source, often mentioned in the same context as HD~100546.  \citet{mee01} suggested that the SED for group Ia sources could be explained by a geometrically thin, optically thick disk responsible for the far-infrared (FIR) emission, as well as a flared optically thin component responsible for mid-IR emission. However, they did not develop quantitative models to confirm if this configuration could indeed reproduce the SED. More recently, \citet{vin06} concluded from the surface brightness profiles derived from the STIS coronagraphic observations \citep{gra01} that the configuration of the material close to the star is that of a geometrically thin, optically thick disk surrounded by a flat, geometrically thick, optically thin envelope. \citet{vin06} used this model to successfully reproduce the near-IR SED ($<6 \mu$m).

For HD~100546, the 3 $\mu$m excess is even smaller and the mid-IR excess larger than in other Herbig Ae/Be stars \citep{bou03}. This is significant because, historically, the source of the (generally) strong 3 $\mu$m ``bump'' in Herbig Ae/Be stars has been difficult to explain, as it is not obviously produced by an optically thick disk. For example, \citet*{har93} have suggested that it is due to emission by an infalling envelope. Recently, \citet*{dul01} showed that the puffed-up inner edge of a Chiang-Goldreich disk (a flared disk with well-mixed dust and gas and a superheated atmosphere, see \citealt{chi97}) subjected to direct irradiation from the star will emit in the near-IR. If this inner rim is the source of the 3~$\mu$m bump, the low 3~$\mu$m emission from HD~100546 suggests a modest rim (outside the bounds of the self-consistent model by \citealt{dul01}) and a shallow dust surface density for the inner disk. With a model of these characteristics on a 400 AU disk \citet{dom03} were able to fit the system's SED, although the fit underpredicts the submillimeter flux by almost an order of magnitude. Because of the small 3 $\mu$m excess, \citet{har05} were able to fit the SED with an unmodified Chiang-Goldreich disk (no puffed-up rim), 150 AU in radius. \citet{har05} did not include any submillimeter measurements in their SED model, which may explain some of the difference in radius with the models from \citet{dom03}. 

A completely different model was developed by \citet{bou03}, who fitted the SED by a set of three optically thin spherical shells. The purpose of their model was not to determine the geometry of the emission but the mass distribution of the dust as a function of temperature. They concluded that the low near-IR emission can be explained by a mostly empty region (an inner hole) within 10 AU and speculated that the hole had been created by a protoplanet.  

Evidence for a hole close to the star has also emerged from spectroscopic data. Ultraviolet observations using STIS \citep{gra05} reveal a deficit of molecular hydrogen close to the star, which the authors interpret as a central gas and dust cavity, 13 AU in radius. In addition \citet{ack06} have used the 6300 \AA ~line of [OI] to conclude that there is a 4.3 AU-wide gap in the disk, 6.5 AU away from the star. They argue that this is consistent with the presence of a 20 M$_J$ object at that position.  

The mid-IR spectrum of HD~100546 is similar to that of Comet C/1995 O1 (Hale-Bopp) \citet*{mal98} in that it shows strong emission in crystalline silicates. In this respect it resembles that of the optically thin debris disk around the K0 dwarf HD~69830 \citep{bei05}. Models of the material around HD~100546 suggest a relatively large fraction of small (0.1 $\mu$m) forsterite and large (1.5 $\mu$m) enstatite grains \citep{mal98, van05}. The models by \citet{bou03} indicate that large amorphous silicate grains and small forsterite grains have different temperatures, unlike the case in Hale-Bopp, although they overlap in radial distance from the star. The mineralogical composition of the dust led \citet{bou01} to suggest that thermal annealing may not have been the origin of the forsterite and that it may result from the destruction of a differentiated larger parent body. All these spectroscopic models indicate a highly evolved disk both in terms of particle size and crystalline fraction.

Optical and ultraviolet observations indicate that, in spite of the inner hole, the star is actively accreting and has a stellar wind \citep{vie99, del04}. While hot molecular hydrogen has been detected close to the star \citep{lec03}, the cold gas component has not been detected \citep{lou90,nym92,wil03}. From measurements of the 1.3 mm dust continuum \citet{hen98} concluded that there are 240 M$_\earth$ of dust within a region 11'' $\times$ 11'' in size, which would imply 100$\times$ larger gas mass, if the mixture has ISM proportions.  This is certainly an upper limit as the non-detection of the HCO$^+$ line at 89 GHz by \citet{wil03} suggests that the gas may be strongly depleted.
 
The lack of detectable gas, the presence of large particles, and the evidence for an inner hole suggest that HD~100546 is an evolved pre-main-sequence star, a ``transitional'' disk perhaps on the way to becoming a more collisionally evolved object like HD~141569 \citep{cla03} or $\beta$ Pictoris \citep{gol06}. In this, it joins a class of objects with signs of evolution in their inner disk (flux deficits in the near IR relative to the median SED of the class) which have been identified in the last decade \citep{nat06}. However, the exceptionally high fraction of crystalline silicates and the characteristics of its SED suggest that it may be in a class by itself.

Because of its relevance in this poorly understood transitional phase, we targeted it for observation using the Advanced Camera for Surveys (ACS; \citealt{for03}, \citealt{pav06}) on board the Hubble Space Telescope (HST). In this paper, we present F435W, F606W, and F814W coronagraphic images of HD~100546's circumstellar environment. These images reveal the environment between 160 and 1300 AU from the star with unprecedented spatial resolution, suppression of the wings of the point spread function, and photometric precision. In addition, they allow us to measure for the first time the color of the circumstellar dust. The scattered-light observations presented here sample the small dust population, far ($\gtrsim$160 AU) from the star. They are then complementary to spectroscopic observations and SED modeling that focus primarily on the inner disk.

%The mass of dust in the disk in these models is $\sim20 M_{\earth}$. 
%Infrared excess implies an optically thick disk, according to Lada.
%Circumstellar material has also been resolved in the thermal infrared, using nulling interferometry and direct imaging. The emission has been modeled as an inclined disk with the same PA and inclination as suggested by scattered light studies. \citet{liu03} concludes that the emission comes from within 20 AU for the mid-IR wavelengths they consider. This is unlike the measurement by \citet{gra01} at 11.7 $\mu$m which suggests that emission at this wavelength comes from within a region 75 AU in radius. 
%Henning et al. (1998): Mg=7.2d-2 Msun=24000 Meaths, 240 Mearths of dust!
%Dominik et al. (2003): Mdisk=5d-3 msun (Not especfied, but it has to be total) Mdust=5d-5 Msun ~ 15 Mearths in the disk.
%Natta et al. (1997): 32d-5 Msun dust ~ 100 Mearths of dust.

\section{Observations and Processing\label{obscons}}

In this section we present a summary of the reduction procedure, which involves the subtraction of a point spread function (PSF) from the target star. The reduction follows the standard procedures common to other coronagraphic observations. The resulting images were then deconvolved using a Lucy-Richardson algorithm. Those deconvolved images form the core data used for the analysis. The science-quality images on which most of the conclusions in this paper are based are available from the Journal's Web site.

These observations were conducted as part of the guaranteed observing time awarded to the ACS Investigation Definition Team (proposals 9987 and 9295). They were performed using the coronagraphic mode of the ACS High-Resolution Channel (HRC), with the 0''.9 occulting spot, which suppresses the stellar PSF wings by factors of the order of ten \citep{pav06}. It is possible to improve the contrast by subtracting a star of the same spectral type as the target, which suppresses the remaining PSF wings. To do this subtraction, we observed  HD~129433 as a PSF reference star. The star was chosen because it is bright, has similar color (V=5.73, B-V=-0.01, \citealp{cra63}). 

The observing log is shown on table \ref{obs_log}.  The images were taken with the F435W (Johnson B), F606W (broad V) and the F814W (broad I) filters. The HRC has a pixel scale of $25$ mas pixel$^{-1}$, and a coronagraphic field PSF FWHM of 50 mas in the F435W passband, 63 mas in the F606W passband, and 72 mas in then F814W passband. For each band and each star, a short direct exposure was followed by a coronagraphic exposure. In each band, HD~100546 was observed at two different telescope roll angles. The process of PSF subtraction produces image artifacts and the two orientations help distinguish between these and real features. 

\subsection{Reduction}

The initial stages of the image reduction (i.e., subtraction of bias and dark frames, and division by a non-coronagraphic flat field) were performed by the ACS image calibration pipeline at the Space Telescope Science Institute (STScI) \citep{pav06}. We started our own image reduction procedures on the separate flat-fielded images before combination, cosmic-ray rejection, and correction of the HRC's geometric distortion by the STScI data pipeline. Coronagraphic images with the same exposure time were averaged and cosmic rays were rejected using conventional IRAF tools\footnote{IRAF is distributed by the National Optical Astronomy Observatories, which are operated by the Association of Universities for Research in Astronomy, Inc., under cooperative agreement with the National Science Foundation.}. To account for vignetting around the occulting spot, the coronagraphic images were manually divided by the spot flats available from the ACS reference files page at STScI's Web site. Each spot flat was shifted to the appropriate position corresponding to the time of observation. This step is done automatically since the CALACS pipeline version 4.5.0.

The coronagraphic images obtained in each band and roll position were then sky-subtracted. The sky was measured by taking the median in the number of counts from circles 20 pixels in radius at the four corners of the images. The effect of this step is minimal in the overall photometry. After sky subtraction, the coronagraphic images were divided by the exposure time. In the longest exposures, light diffracted by the coronagraphic mask saturates some pixels in the region within 2'' from the star and we replaced those saturated pixels by non-saturated pixels from shorter exposures. 

\subsection{A Note on HST's Photometric systems \label{hst_phot}}

The F435W filter is a good approximation of Johnson B, but the other two filters used in these observations do not correspond to standard filters. To calibrate our observations and compare them with published results, we make use of three different photometric systems: an ``instrumental'' vegamag system, a standard vegamag system, and the standard Johnson-Cousins system as described by \citet{lan92}.

The instrumental vegamag system is described by \citet{sir05}. In this system Vega has zero magnitudes and zero colors in all bands. The transformation from e$^{-}$/sec to magnitudes involves only a zero-point and no color term. We will refer to magnitudes in this system with the filter name. For example, the F435W magnitude, etc. This is the system most used in this paper.

If the spectrum of the source is known, it is possible to obtain magnitudes in the standard vegamag system. These magnitudes are given in the Johnson B, Johnson V, and Cousins I bandpasses, but the calibration is still tied to Vega. Transformations between the instrumental and the standard vegamag system are possible using the synthetic photometry package SYNPHOT, which is distributed by STScI. We will refer to the magnitudes in this system with the subscript {\it Vega}. For example, B$_{\rm Vega}$, V$_{\rm Vega}$, etc.

Ground based observations usually present measurements in a standard photometric system, involving standard filters and a grid of primary calibrators. \citet{sir05} provide color transformations between the instrumental vegamag system and the standard system described by \citet{lan92}. \citet{sir05} argue that the transformations are strongly dependent on the details of the source spectra and so we use them sparingly.

\subsection{Stellar Photometry \label{stell_phot}}

A measurement of the stellar flux is required for two aspects of the analysis. First, in order to subtract the PSF reference star from the target, the former needs to be scaled by a multiplicative factor. This factor is the ratio of the brightness of the reference star and that of HD 100546 in each band. Very accurate photometry is not crucial for this procedure because visual inspection of the subtracted images is enough to determine the scaling factor within $\pm3$\%.  Second, in order to determine the intrinsic color of the circumstellar material, we divide the images by the stellar flux. For this, we need to know the stellar photometry as accurately as possible. As one of the main conclusions of this paper is that the circumstellar material is intrinsically red, we now describe in detail the procedure followed to obtain this photometry. 

For F435W we used the direct (non-coronagraphic) images of HD~100546 and HD~129433 to measure the number of e$^{-}$/sec within 5''. While the cores of the stellar PSFs are saturated, observations performed with the parameter GAIN set to 4 (as are these) preserve the number of electrons over the whole image \citep{gil04}. In other words, as long as the aperture is large enough to encompass all the saturated pixels, the integrated number of counts can be recovered.  

Direct non-coronagraphic observations in the other bands were executed with GAIN=2 and so the number of counts in saturated images is not conserved. We used the number of counts measured in F435W, a model spectrum of the target and the PSF reference star, and the measured colors of HD~100546, to predict the number of counts in F606W and F814W.

For HD~100546, \citet{dew01} measured B-V=$0.012\pm0.005$ mags and V-I=$0.042\pm0.01$ mags, from which \citet{van98} derived A$_V$=0.28 mags, assuming that the intrinsic color was  B-V=-0.075 mags. In order to compare with our data, we assumed that the colors in the standard vegamag system are the same as those in the standard Johnson-Cousins system. This is a very good assumption for HD~100546, whose spectral type is B9.5 and so the color-dependent terms in the transformation equations are small \citep{sir05}. As the model spectrum for the target we used the B9V star HD~189689, from the Bruzual-Persson-Gunn-Stryker (BPGS) Atlas provided by STScI. This star has B$_{\rm Vega}$-V$_{\rm Vega}$=-0.11. We reddened the spectrum using the standard extinction law \citep*{car89} until B$_{\rm Vega}$-V$_{\rm Vega}$=0.012 mags, the average measured color for HD~100546. The resulting extinction is A$_V$=0.36 mags. With this spectral model and the number of counts measured in the F435W band, we used the {\it calcphot} routine within SYNPHOT to derive a B$_{\rm Vega}$ magnitude. The B$_{\rm Vega}$ magnitude derived in this way (table \ref{photom}) is 0.03 mags less than the average B magnitude derived by \citet{dew01}, consistent the difference  between the standard system, in which Vega has B=0.03 mags, and the vegamag system, for which Vega has B$_{\rm Vega}$=0.00 mags. From the stellar colors of HD~100546 given by \citet{dew01} we derived V$_{\rm Vega}$ and I$_{\rm Vega}$. With these magnitudes, the {\it calcphot} routine, and the same spectral model, we obtained the number of counts in F606W and F814W. The result of this procedure is that the number of counts in each band corresponds to the measured stellar color. 

The BPGS Atlas contains four more dwarf stars with spectral types between B9 and A0. For each of them, a different value of A$_V$ is required to reproduce the observed color for HD~100546. They range from A$_V$=-0.077 mags in the case of 58 Aql to A$_V$=0.18 mags for $\theta$ Vir (both classified as A0V in the BPGS Atlas). However, following the same procedure described above results in changes of the number of counts for each band of $<$1\% with respect to those reported in table \ref{photom}. The procedure is robust to the assumed spectral model because each model is forced to have to observed colors of HD~100546, independently of its intrinsic colors.

For HD~129433, the number of counts in F606W and F814W was obtained from the number of counts in F435W by using as spectral model the unreddened spectrum of HD~189689. The PSF reference star has a larger B$_{\rm Vega}$-V$_{\rm Vega}$ (by 0.1 mags) than HD~189689, and so this procedure will systematically underestimate its flux by $\approx 10$\%. Accurate photometry of the reference star is not crucial for the conclusions of this paper.

\citet{sir05} indicates that the best aperture photometry using the HRC will have 3\% errors. For HD~100546, the color measurements from \citet{dew01} introduce an additional 2\% error (3\% error) in the determination on the F606W (F814W) flux. For the PSF reference star, we assume that the errors are those of the aperture photometry only. These are the errors quoted in table \ref{photom}. 
 
\subsection{Subtraction of the Coronagraphic PSF}
\label{subtraction}

For each band, the normalized, cleaned, and combined coronagraphic images of the PSF reference star HD~129433 were subtracted from each roll of the combined images of HD~100546. The positions of the occulted stars were determined within one pixel using the central peaks of the coronagraphic PSFs that result from the re-imaging of the incompletely occulted, spherically aberrated starlight \citep{kri03}. Further alignment was done visually, by shifting the PSF reference images until subtraction residuals were minimized. The uncertainty in the registration of the images along each axis is 0.125 pixels. For the normalization constants we performed the following visual, iterative adjustment. Starting with the flux ratios implied by the values from table \ref{photom} (0.391, 0.443, 0.474 for F435W, F606W, and F814W, respectively), we adjusted the scaling until the subtraction residuals were minimized. This occurred for normalization constants 0.383, 0.382, and 0.387 in F435W, F606W, and F814W. Visually, the uncertainty of these normalization values is 3\% and in what follows we will propagate this error linearly (as a systematic error) to estimate uncertainties in calculated quantities (Section \ref{sbps}). The values of the normalization constants are consistently lower than implied by the values quoted in table \ref{photom}, as expected by the mismatch between the colors of the PSF reference star HD~129433 and those of its spectral model. 

\begin{figure}
\plotone{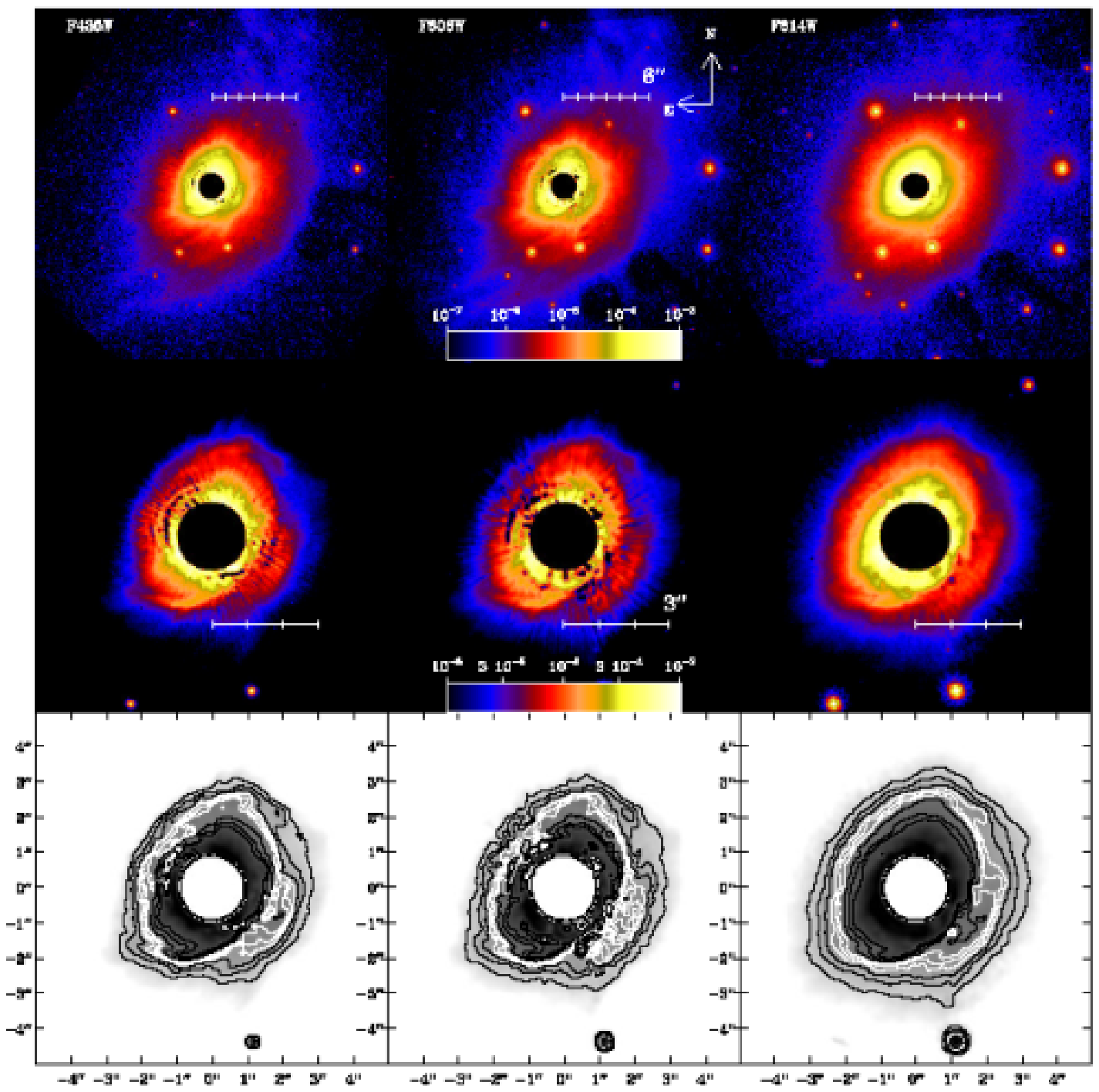}
\caption{\label{panel1}Surface brightness maps of the circumstellar environment of HD~100546, in a logarithmic stretch. All images have been normalized to the stellar brightness in their respective bands. The approximate size of the coronagraphic mask is shown as a black circle  1'' in radius. The top row has a different color stretch and spatial scale than the other two, in order to showcase different regions. The top stretch goes from  $10^{-7}$ arcsec$^{-2}$ to $10^{-3}$ arcsec$^{-2}$. The middle and bottom rows show a stretch from  $10^{-5}$ arcsec$^{-2}$ to $10^{-3}$ arcsec$^{-2}$. The contours in the bottom row are obtained from images that have been heavily smoothed. The contour values are 2, 3, 4, 5, 6, 7, 9, 12, and 15 $\times 10^{-5}$ arcsec$^{-2}$.}
\end{figure}

After subtracting the coronagraphic images of HD~129433 from each image of HD~100546, we corrected for the geometric distortion in the HRC image plane, using the coefficients of the biquartic-polynomial distortion map provided by STScI \citep{meu02} and cubic convolution interpolation to conserve the imaged flux. By using standard IRAF routines, we aligned the two rolls in each band, using other stars in the field for reference, and averaged the images from each roll. The resulting surface brightness maps (divided by the stellar fluxes) are shown in figure \ref{panel1}. These maps are available from the Journal Web page.

The mid and bottom rows of figure  \ref{panel1} highlight different regions of the circumstellar environment. Differences in the shape of the PSF between HD~100546 and the reference star HD~129433 (due to differences in color, telescope focus, or position behind the mask) result in subtraction residuals. The radial streaks in the top two rows in figure \ref{panel1} are produced by differences in the way light is scattered by telescope optics for the two stars while the concentric dark and bright rings are due to differences in the diffraction pattern produced by the coronagraphic mask \citep{kri01}. The rings are more noticeable along the NE-SW direction, which coincides with the position of the scattering ``strip''. This well-known artifact of the ACS coronagraph is a bar-like region of increased number of counts, centered on the star and $\approx$6'' long (e.g. \citealt{kri05a}). 

The presence of these residuals makes the analysis within $\approx 1.6$'' of the star impossible. Beyond this limit we can use the images in the two separate telescope rolls to distinguish subtraction residuals from features associated with the material around HD~100546, as the residuals are nearly fixed on the detector's axes. To trace the contours shown in the bottom row of figure \ref{panel1} we perform a 10-pixel median azimuthal smoothing on the middle-panel images, followed by a 5-pixel boxcar average smoothing. 

In figure \ref{panel1}, the difference in the spatial resolution among the three bands is seen as a smoothing of the images with increasing wavelength. In the top row, the imprints of the larger coronagraphic spot and the occulting bar are visible towards the SW direction (one for each roll angle).

\subsection{Red Halo and Image Deconvolution}

For HRC observations in the F814W band, one has to contend with the effect of the red halo \citep{sir05}. This may appear as a diffuse, smooth halo of light, slightly asymmetric but fixed with respect to the detectors axes. The halo is composed of photons from the core of the stellar PSF which have been scattered to large angles by the CCD substrate. The halo asymmetry can be seen in the direct (non-coronagraphic) images from \citet{kri05a} (See their figure 2). In this section we show that deconvolution of the stellar PSF, which partially corrects this effect, changes modestly the slope of the surface brightness profiles. 

To deconvolve the images, we use the PSFs generated by the Tiny Tim software package, distributed by STScI \citep{kri04}. The simulated ``off-spot'' coronagraphic PSFs can be generated for an array of spectral types.  These PSFs include a simplified model of the red halo but do not include the red halo asymmetries mentioned before. We do not believe those asymmetries are important in our images: upon subtraction of the two rolls no residual asymmetric structures are revealed. Unlike the non-coronagraphic images from \citet{kri05a}, the coronagraphic images presented here do not produce a very strong core to be scattered at large angular distances. 

The deconvolution is performed on the image after subtraction of the reference PSF. Therefore, only
the occulted star subtraction residuals contribute to the deconvolution process (and then basically as noise) as the effects of vignetting near the occulter are removed with the division by the coronagraphic spot flat field. The only effect of the coronagraphic setup on emission outside of the focal plane mask is to alter the PSF, which is then defined by the Lyot stop rather than the entrance aperture of the telescope. Simulations \citep{whi07} confirm that beyond 1.5'' from the center of the coronagraphic mask, the off-spot PSF does not depend on distance. 

\begin{figure}
\epsscale{1}
\plotone{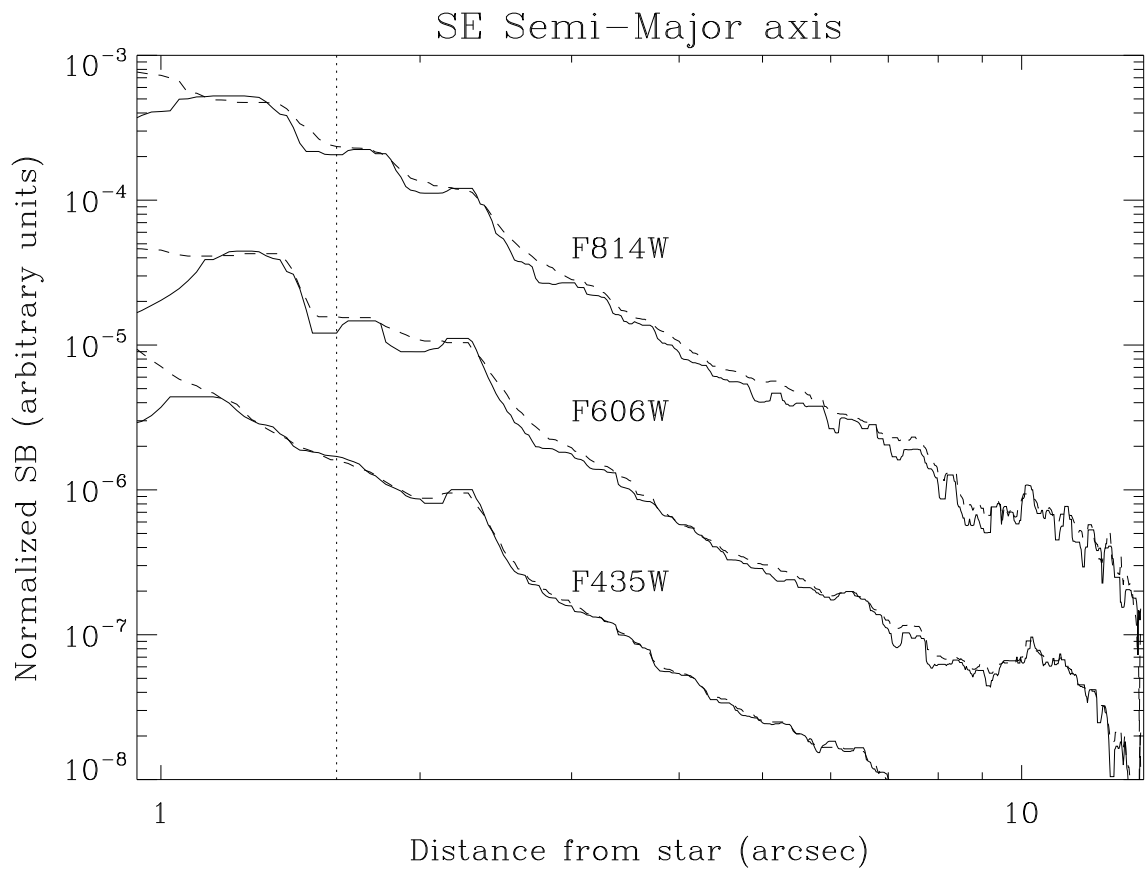}
\caption{\label{before_after} The effect of the deconvolution in the surface brightness profiles. The plot shows the median surface brightness in segments 0.''25 long, taken along the SE extension of the major axis of the disk (145 degrees East of North). Each profile has been smoothed by a median boxcar 10 pixels wide, and offset by an arbitrary amount in the y-axis. The dashed lines show the profiles before, and the solid lines show the profiles after the deconvolution. The dotted line is placed at 1.''6 from the star, the distance within which the images are dominated by subtraction residuals. Note that the effect of the deconvolution is to sharpen the features within $\sim$2.''5 At larger distances, the deconvolution is most important for the F814W band, resulting in a surface brightness profile reduced by $\sim$15\% with respect to the non-deconvolved one.} %updated to new decon
\end{figure}

\begin{figure}
\plotone{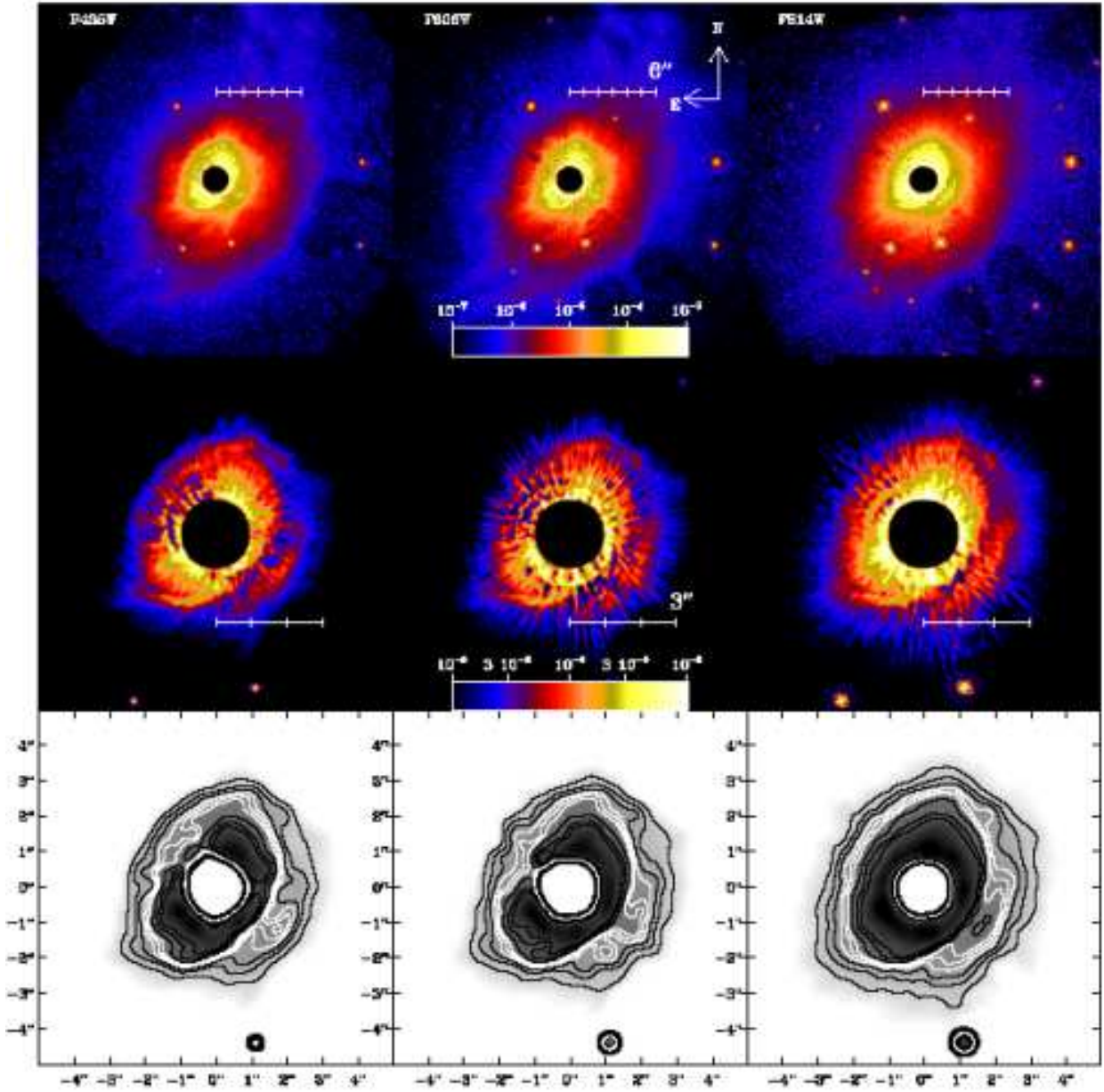}
\caption{\label{panel_dec} PSF-deconvolved images of the circumstellar environment of HD 100546.  The 
color and spatial scales are the same as shown in Figure \ref{panel1}.  In order to draw the contours, the images have been smoothed like those in the bottom row of figure 1, and then convolved with a 10-pixel-FWHM 
gaussian (see Section 2.6).} %updated
\end{figure}

We generated distorted off-spot PSFs in each band, using BPGS model of the A0 star $\theta$ Vir as our model. This is the closest spectrum to our target provided by Tiny Tim. We applied the Lucy-Richardson deconvolution algorithm \citep{luc74,ric72} as implemented in IRAF to each (single-roll, distorted, PSF-subtracted) image outside a circular region 1'' in radius, centered on HD~100546. We terminated the computations at 50 iterations. By this point, we could not detect changes in the FWHM of the field stars, nor in the surface brightness profiles of the image. After deconvolution, we corrected for camera distortion, rotated the two rolls in each band to a common orientation and averaged them. In this way, we obtained a deconvolved image in each band.  A comparison of the ``before'' and ``after'' surface brightness profiles in each band is shown in figure \ref{before_after}. The resulting images are shown in figure \ref{panel_dec}.

Figure \ref{before_after} shows that the deconvolution has its strongest effect in the F814W images, where it results in a sharpening of features within 2.''5 and a modest reduction of the large scale surface brightness from 2.''5 to 10'' away of HD~100546. As mentioned before, the deconvolution process is not valid within 1.''5 from the coronagraphic mask. In the case of the direction shown in figure \ref{before_after} (corresponding to the SE semi-major disk axis), the deconvolved profiles are $\sim$15\%, $\sim$5\%, and $\sim$3\% fainter in average than the non-deconvolved ones, for F814W, F606W, and F435W, respectively. These values are similar to those along other directions. Therefore, the $\Delta (F435W-F814W)$ color of the deconvolved images (defined in section \ref{sbps}) is $\sim$0.12 mags less (i.e. not as red) than the color of the non-deconvolved ones. As we will show in section \ref{disk_color} this is comparable to the errors in the color. So in terms of the large scale behavior of the color it does not make much difference if the analysis is carried out in the deconvolved or in the non-deconvolved images. 

Figure \ref{panel_dec} shows the deconvolved equivalent of figure \ref{panel1}. After deconvolution, the features in the circumstellar environment become sharper, and the images become noisier. The amplified, correlated noise in the deconvolved images is characteristic of the Lucy-Richardson algorithm and it is a consequence of the algorithm's requirements of a low-background signal, positive pixel values and flux conservation in local and global scales. The higher noise levels require that the images be further smoothed before calculating the contours. To measure them, we smooth the images as in figure \ref{panel1} and then convolve them with a gaussian function having FWHM=10 pixels, which acts as a low-pass filter. It is in these heavily smoothed figures that the contours are traced.

The deconvolved F435W, F606W, and F814W surface brightness maps are available from the Journal Web page.

\subsection{Color-combined images}

In order to grasp the overall appearance of the system, we created color composites of the images by mapping the bands in three color channels: F435W in Blue, F606W in Green and F814W in Red. Before combining them, each image was divided by the stellar flux in its band. To produce the color composites, the images were scaled by the inverse hyperbolic sine of the average in all three bands, using the softening parameter $\beta=0.005$, as described in \citet{lup04}. This scaling makes the apparent color independent of the brightness. In effect, the scaling is linear for very faint regions and logarithmic for brighter ones. 

\begin{figure}
\epsscale{1}
\plottwo{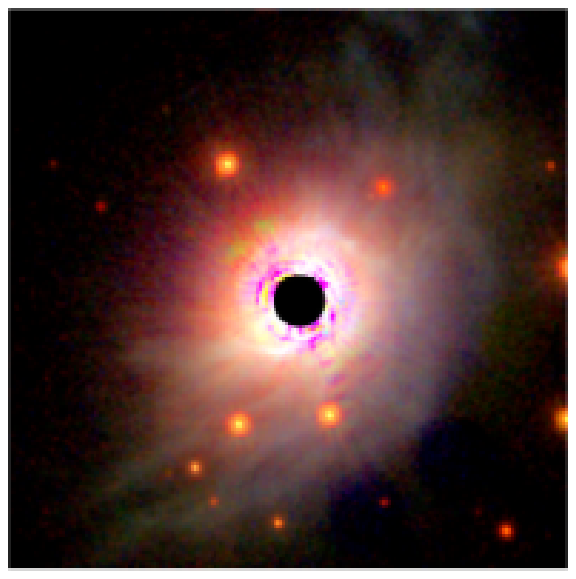}{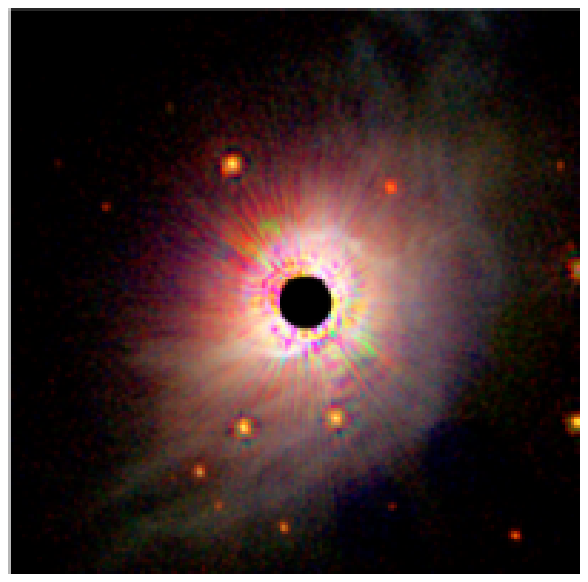}
\caption{\label{combi}RGB composite of the individual images. As in previous figures, the black central circle is 1'' in radius. North is up and East is left. Before combining them, each image has been normalized by the stellar brightness. In order to map the number of counts to the RGB channels, we have used the algorithm described by \citet{lup04}. Left: Color combined image. Right: Color combination of the deconvolved images.}%(MISSING: COMPASS ROSE \& SCALE) 
\end{figure}

The results are shown in figure \ref{combi}. Regions of extreme colors, like the circular purple features close to the coronagraphic mask, indicate places in the image for which data from at least one of the bands are missing. These occur because the position of the coronagraphic subtraction residuals is wavelength dependent. The deconvolution process (figure \ref{combi}, right) makes the radial streaks resulting from PSF mismatch very evident.

\section{Results\label{results}}

Figures \ref{panel1}, \ref{panel_dec}, \ref{combi}, show the well known circumstellar structure of HD~100546 \citep{gra01} but with an unprecedented degree of clarity. Within 1.''6 the images are dominated by subtraction residuals. From 1.''6 to 10'' from the star, the total scattered-light relative fluxes, measured in the deconvolved images, are $(L_{Circum}/L_{*})_{\lambda}=1.3\pm0.2\ \times 10^{-3}$, $1.5\pm0.2\ \times 10^{-3}$, and $2.2\pm0.2\ \times 10^{-3}$, in the F435W, F606W, and F814W bands respectively. These values are $\approx 200-400$ times less than the total IR excess.  

In this section we describe the main observational results associated with the general characteristics of the scattered light distribution around HD~100546. 

\subsection{Photometry of Field Stars}

HD~100546 has a galactic latitude of -8.32 degrees, which suggests that the other point sources in the images are unassociated field stars. In table \ref{field} we list the standard photometry for the brightest stars within $\sim$6'' of HD~100546. See also figure \ref{photo_stars}. We only list those objects that are observed in both rolls (and are outside the occulting bar in both) in at least one band. To compare with known stellar colors, the photometry is given in the standard Johnson-Cousins system, using the color corrections from \citet{sir05}. 

The largest V-I color is $\sim2$. If the object is unreddened, this color would suggest that it is an early M dwarf. Such a star would have an I magnitude of $\approx 12$, much brighter than observed \citep{leg92}. All other objects have bluer colors, indicating earlier spectral types and brighter predicted apparent magnitudes. Therefore, all objects in table \ref{field} are farther away than HD~100546.  

\subsection{General Morphology}

Overall, the images show a broad band of light wrapping around the SW side of the star (figure \ref{combi}, top), reminiscent of the configuration of the scattered light images around AB Aurigae \citep{gra99}. A red glow can be seen on the NE side of the target both in the deconvolved and non-deconvolved images, similar to the strong F160W scattered light signal found closer to the star by \citet{aug01}. 

Interpreting the inner circumstellar surface brightness as due to an inclined disk, we use the deconvolved images to fit elliptical isophotes to the images, with semi-major axes ranging from $1.''6$ to $\sim 2''$. At distances larger than 2'' the contours become ``boxy'' because of the bright SW feature (structure 2, see section \ref{spiral}) and the scattered light cannot be fit with an ellipse. For all the bands, the position angle (PA) of the semi-major axis is $\approx 145$ degrees. From the ellipticity of the isophotes we derive an inclination angle of 42 degrees from face-on. There are no significative differences in these values among the three bands. This is an observational determination of the inclination and PA based on isophotal contours, and ignores the effect of anisotropic scattering of the stellar light by the dust grains. These results are summarized in table \ref{geometry}.  

Table \ref{geometry} also compares the disk geometry derived in this work with other published results. All disk inclinations are consistent, but the position angles of the major axes are only marginally so. Our results differ from those by \citet{pan00}, \citet{gra01}, and \citet{aug01} at the 2$\sigma$ level. \citet{pan00} derived the PA by using symmetrized image of the disk. \citet{gra01} do not describe the procedure used to determine the PA. \citet{aug01} determined the PA by performing isophot fitting within 3''. These differences may reflect the fact that a measurement of the ratio of the axes of an ellipse is less sensitive to uncertainties than measurements of a single axis. Furthermore, there may be differences between the value of the scattering phase function in the optical and the near-IR. 

For the F814W images, the contours within 2'' (bottom row of figures \ref{panel1} and \ref{panel_dec}) are closer together at the SW side than at the NE side: the image brightness changes faster in the former than in the later. The contours in the other bands are not as well defined at these distances. 

In the sections that follow and for the purposes of the morphological description, we divide the circumstellar environment in three regions, where the distances are measured along the disk major axis: the inner disk (from 1.''6 to 3''), the mid-disk (from 3'' to 8'') and the extended envelope (beyond 8''). Each region has its own dynamical and/or radiative characteristics. The inner disk and the material under the coronagraphic mask, comprise the region believed to be responsible for most of the far-IR excess.
 
\subsection{Structures in the Disk}
\label{spiral}

The well-known ``spiral arms'' of HD~100546 (e.g.  \citealt{gra01}) are clearly seen beyond 1.''5 to $\approx$3'' from the star, as elongated structures at the NW, SW, and SE (figure \ref{anot_struct}). The simpler mask shape of the ACS/HRC coronagraph compared to the STIS one, allows for a more complete description of the environment. In the rest of the paper, we refer to these features as structures or spiral arms, interchangeably. 

Structure 1 is a narrow band of brightness visible between 2.''3 and  2.''8 from the star, and figure \ref{anot_struct} suggests that it may continue on the NE side of the disk. Structure 2 describes a broad arc at 2'', and is separated from the inner part of the emission by a dark region $\sim $0.''5 wide. However, the exact extent of the dark lane is hard to judge because of the subtraction residuals along the minor axis. Structures 1 and 2 are separated by a region that is half as bright as the maximum of structure 2. Structure 3 extends from 1.''5 to 2.''5 from the star, and it seems more open than structure 1. Because of this, structure 1 seems to wrap itself around the NE side of the disk, while structure 3 stabs the SW side. Structures 1 and 3 are superficially reminiscent of galactic spiral arms. The morphology of structure 2 is reminiscent of that produced when starlight is scattered from the disk back side towards the observer (see for example, the observations of GM Aur in \citealt{sch03}). However, both in observations and simulations \citep{whi92}, such structure is generally narrower and it wraps around the disk over a larger angle than seen here. 

Observations of the [OI] 6300 \AA ~line by \citet{ack06}, reveal blueshifted emission from the SE of the disk and redshifted emission from the NW part, suggesting that the inner part of the disk is rotating counter-clockwise. Their observations trace gas at distances $<$100 AU from the star. Assuming that the disk rotates in the same direction at larger distances from the star and that the SW side is oriented towards the Earth (section \ref{morpho}), those spectral observations indicate that the structures are trailing the direction of rotation.

Before deconvolution, the behavior of the structures with wavelength is tangled with the changing resolution of the telescope at different passbands (figure \ref{panel1}). The deconvolved images (figure \ref{panel_dec}) reveal no obvious morphological differences in each band, for each of the structures. The contour levels traced on the deconvolved images show that the space between the inner part of the disk and the structures becomes brighter at longer wavelengths, while the value of the peak brightness does not change much between bands. This color behavior is discussed in the section \ref{sbps}. 

The circumstellar disk is illuminated by the starlight, which decays as the inverse distance squared from the star. An appropriate correction for this effect is only possible for face-on disks. We therefore deproject the disk by the inclination, divide by the Henyey-Greenstein scattering phase function with $g=0.15$ (derived in section \ref{morpho}), and multiply every pixel by the square of the projected distance to the star. This procedure will reveal the correct geometry and brightness of the circumstellar material only in the case of a disk in which every dust particle is illuminated by the full stellar radiation field. Such is not the case for HD~100546, as every reasonable model of the SED indicates that some part of the disk is optically thick. However, independently of the optical depth, the procedure helps to reveal weak disk features and clarifies the disk structure, although the brightness of the structures and their distance to the star, particularly along the minor axis, will be incorrect. 

The result is shown in the top left panel of figure \ref{mosaic435} for the non-deconvolved images in the F435W band (images obtained from other bands and from the deconvolved data look similar). Because the hemispheric subtraction residuals are circular in the original image, they appear elliptical in the deprojected one. In order to accentuate the sharp structures, we subtract from the deprojected image a smoothed version of itself (figure \ref{mosaic435} top right). The smoothed image has been convolved with a gaussian function having a 30-pixel FWHM. The subtraction of this smoothed image from the deprojected one leaves only the sharp features. An unsharp masking technique similar to this has recently been used to clarify the arm structure of our galaxy \citep*{lev06}. 

This unsharp masking  also sharpens the subtraction residuals present in the images, and so identification of real features requires care. In the bottom right panel of figure \ref{mosaic435} we have marked those features that are observed in all three bands. The original structures 1, 2, and 3 are traced with solid lines, and marked as 1a, 2a, and 3a, while new structures are traced with thick dashed lines. 

The brightest new features are structure 1c, an inner arm at $\sim2$'' that seems to emanate from 1a, and structure 1d, a feature behind the disk. The deprojected images show that structure 2a curves toward the star, like another spiral arm. The 1a+1b arm is $\sim$5'' long, while 2a+2b and 3a+3b are $\sim$4'' long. On the deprojected, unsharp masked image, arms 1a and 3a have the same opening angle with respect to the disk and appear like symmetric images of each other. Together with arms 2a and 1d, the overall configuration is that of a four-armed spiral. The unsharp masking traces only the sharp ridges, and so the features may extend beyond these limits.

Structures analogous to these have also been detected around AB Aur, based on ground-based coronagraphic images in the H-band \citet{fuk04}, although they do not seem as well ordered. In those observations, at least one of the arms appears split, as structure 1 is split between 1a and 1c. The arms around AB Aur are also believed to be trailing the disk. 
 
\subsection{Surface Brightness Profiles\label{sbps}}

Figures \ref{major} and \ref{minor} show the relative surface brightness profiles (SBPs)  and colors, along the major and minor axes of the disk. The SBPs are obtained from the deconvolved images by taking median values along rectangular strips (0.''25 wide) centered on the star for each band. In addition, each profile has been median-smoothed with a boxcar function 10 pixels long. The bottom plots show the colors measured with respect to the star: $\Delta (F435W-F606W)=-2.5 \log (SBP_{F435W}/SBP_{F606W})$, $\Delta (F435W-F814W)=-2.5 \log (SBP_{F435W}/SBP_{F814W})$. An aid to the interpretation of the profiles is provided in figure \ref{circles}. 

%Figures \ref{colors} show the  $\Delta (F435W-F606W)$ and $\Delta (F435W-F814W)$ colors. Structures 1 and 2 produce ``bumps'' in the SBPs and dips in the color profiles. Structure 3 is not noticeable in the scale of the plots. 

\subsubsection{Sources of Error and Comparison with Previous Measurements}

There are three main sources of error in the SBPs: photometric errors due to the uncertainty in the stellar flux (which varies between 3 and 5\%, as given in table \ref{photom} and result in errors small compared to those from other sources), photometric errors due to the uncertainty in the flux ratio between HD~100546 and HD~129433, and errors (variability) within the 0.''25 segments.  In figures \ref{major} and \ref{minor} the total 1$\sigma$ errors are shown for some representative points. For these profiles, the 3$\sigma$ detection limits are $2\times 10^{-7}/arcsec^2$, corresponding to V$_{\rm Vega}$=23.4 magnitudes/arcsec$^2$.

The uncertainty in the flux ratio produces errors which scale with the brightness of the circumstellar material. We propagate them linearly, as they are systematic and not random. For the major axis, they range from 10\% for distances $<$8'' to 5\% farther out. For the minor axis they range from 20 to 40\%. The errors are larger along the minor axis because that axis coincides with the position of the scattering strip (see section \ref{subtraction}). Finally, errors in the median within the 0.''25 segments are a measure of the variability within the strips used for the calculation of the SBPs, and range from $\approx$5\%, 2'' from the star, to $\approx$10\% at 12'' for the major axis or $\approx$40\% at 8'' for the minor one. 

In summary, errors in the photometry (introduced by errors in the normalization constant) dominate the error budget at small distances and uncertainties due to variability within the median strip dominate at large distances, with a turnover point at 6-8'' from the star. 

In addition, there are regions of the profiles strongly affected by the systematic mismatches between  HD~100546 and the PSF reference star HD~129433. For example, the ``dip'' in the SE semi-major axis in F606W at 1.''5 is due to over-subtraction of the PSF at that position, as it is obvious from figure \ref{panel_dec}. Within $\approx 1.''6$ the SBPs in all directions are dominated by subtraction residuals.

As mentioned in the introduction, previous coronagraphic observations of the circumstellar environment are available. Over the unfiltered STIS passband, \citet{gra01} measured a total count rate of 126 cts/sec at 2'', on 9 pixels, which implies 5382 e$^{-}$/sec/arcsec$^2$.  With this measurement, the reddened spectrum of HD~189689 from section \ref{stell_phot}, and {\it calcphot}, we predict that the count rate in the F606W passband should be $5\times 10^{3}$ e$^{-}$/sec/arcsec$^2$, or $1\times 10^{-4}$  per arcsec$^2$ (after dividing by the stellar flux in the F606W passband). This point is shown in figure \ref{major}. We repeat the analysis for a point at 8''. Both points are broadly consistent with our observations. To compare with the NICMOS coronagraphic images we use the fact that \citet{aug01} measured 16 mag/arcsec$^2$ at 2.''5 from the star, which implies 1.7$\times 10^{-4}$ /arcsec$^2$, at the F160W band. To obtain this number we used the magnitude zero-points reported by  \citet{aug01} and the reddened spectrum of HD~189689 as a spectral model, which implies that the F160W flux density of HD~100546 is 2.5 Jy. We also plot the NICMOS measurement at 3.''5 for which \citet{aug01} measured 18 mag/arcsec$^2$. The NICMOS measurements imply $\Delta (F435W-F160W)=$1.5 at 2.''5 and $\Delta (F435W-F160W)=$1.4 at 3.''5. As before, uncertainties in the stellar model translate into uncertainties of $\approx$ 4\% in the calculated flux. Comparison with the ground-based ADONIS measurements reported by \citet{pan00} yields results that are difficult to understand. They measured 0.3 Jy/arcsec$^2$ at 0.''5 from the star in the J-band. This is one order of magnitude larger than the F160W NICMOS measurement at that point ($\approx 11$ mag/arcsec$^2$ or 0.04 Jy/arcsec$^2$).  \citet{sch04} has shown that variations in the PSF shape due to atmospheric fluctuations may result in false disk detections, when using ADONIS paired with coronography. On the other hand, the PA and inclination reported by \citet{pan00} are roughly consistent with those derived using other instruments.

\subsubsection{Morphology of the Nebulosity}
\label{morpho}

The top panels of figure \ref{major} show that at 8'' the F814W SBP drops by a factor of $\sim$2. The drop is very abrupt along the SE direction. Beyond this distance, the $\Delta (F435W-F814W)$ and $\Delta (F435W-F606W)$ colors become indistinguishable from each other. Consistent with the measured inclination, there is a decrease along the SW minor axis at 6''--7''. This is why we define the mid-disk as the disk between 3'' and 8'' from the star, and the extended envelope as material beyond 8''.

Figure \ref{circles} shows that the nebulosity is slightly more extended to the NW side than to the SE. Along the NW side, the major-axis SBPs show scattered light to the edge of the detector, $\sim14''$.  Along the SE side the emission drops at 12'', from $6\times 10^{-7}/arcsec^2$ to below the detection limit. Again, this is consistent with the decrease along the NE semi-minor axis at $\sim 9''$. In other words, the extended envelope is a flattened nebular structure $\approx$12'' -- 14'' in radius, seen with an inclination similar to the inner disk. 

In figure \ref{reflec} we compare the profiles along each semi-axis to each other. The semi-major axis profiles along each direction are very similar to each other, while the profiles along the semi-minor axis are very different. Along the major axis, the main differences between the two sides are in the presence of structure 1 at 2.''8, structure 3 at 2.''3, and field star \# 8 at 5.5". Along the minor axis, the amount of scattered light in all bands is larger in the SW side than in the NE side, until the sharp decrease of the SW side at 6''--7''. The ratio of the scattered light in the SW side to that in the NE side, in each band, between 2'' and 7'', presents strong local variations, with values oscillating between 1 and 4. The ratio averaged over all bands in this range is 2.3. For larger distances, the ratio at all bands is consistent with one. 

If the brightness asymmetry between the SW and NE sides is interpreted as being the result of forward scattering, it implies that the SW is oriented towards the Earth. Assuming that the scattering phase function can be parameterized as a Henyey-Greenstein function \citep{hen41} the average value of 2.3 for the SBPs ratios implies a scattering asymmetry factor $g\approx0.15$. A ratio of 4 would imply $g\approx0.23$. For comparison, ISM grains have g$\approx0.4$ to 0.6 in the optical \citep{wei01}. In other words, the grains around HD~100546 scatter light more isotropically than grains in the ISM. This determination of $g$ is not strictly correct as the procedure is only valid for optically thin dust distributions.

\subsubsection{Power-law Fits}

In any direction, the SBPs are considerably more complex than any previously published. For the purposes of comparison with previous works, we derive the power-law fits (see table \ref{power}) for the SE Major axis. The profile can be broken in four parts: 1.''6 to 5'', 5'' to 8'', and 8'' to 12''. To do the fit, we interpolated over the signature of structure 3. The results of the fit for F435W are shown in figure \ref{reflec}. 

While it seems clear that the profile to 5'' can be described by a power-law, such description is less warranted beyond this limit. In particular, there seems to be a transition region from 5'' to 8'', before a flatter edge at longer distances. Within the errors, the fits are the same across wavelengths.

Power-law fits to SBPs were derived by \citet{gra01} based on STIS observations, using a different PA for the disk, different width for the evaluation of the SBP, and different distance ranges for the power-laws. In table \ref{power} we also derive the power laws using the \citet{gra01} ranges and tracks, but based on our data. For the exponents, they derive $-3.1\pm0.1$ between 1'' and 2.''7,  $-3.1\pm0.1$ between 2.''7 and 5'', and $-2.2\pm0.2$ between 5'' and 8''. Within the errors, their results are consistent with ours except in the range between 2.''7 and 5'', were our slopes are significantly steeper. The origin of this discrepancy cannot be ascribed to the differing observation bands, as the power-law indices derived in all our filters are alike within the errors. From NICMOS F160W data, \citet{aug01} derive $-2.92\pm0.04$ between 0.''5 and 2.''5 and -5.5$\pm$0.2 from 2.''7 to 3.''7.  Between 2.''7 to 5.''0, our measured index is between those of STIS and NICMOS. The difference with NICMOS may be due to the differing optical depths of the material at different wavelengths. 

Beyond 5'', \citet{gra01} fitted a power-law index of -2.2 along the major axis of the disk. Based on this fit, and the models from \citet{whi93}, they concluded that they had detected an infalling envelope in the star. The applicability of the models by \citet{whi93} presumes that the material is optically thick beyond 500 AU from the star. By using the same position angle and strip size to measure the major axis, we can roughly reproduce the power-law dependence from \citet{gra01}, within the errors (See table \ref{power}). However, at large distances from the star, the material will be optically thin along the line of sight. The index of the power-law then reflects the surface density of scatterers, and not the shape of the scattering surface. As such, the use of the models developed by \citet{whi93} to explain the large scale power-law is unwarranted.

\subsection{Disk Color}
\label{disk_color}

The SBPs show that along the principal axes of the disk the circumstellar material is brightest in F814W than in the other bands. Furthermore along the major axis, $\Delta (F435W-F814W)$ increases from 0.3 at 2'', to $\approx$ 1 at 5'', indicating that the dust is between $\approx$ 1.3 and 2.5 times brighter in the F814W band than in F435W. After 8'' there is a sharp decrease in the value of $\Delta (F435W-F814W)$.  The noise precludes an accurate determination of the value, but at least from 8'' to 12'' it seems positive or neutral. For the other color, $\Delta (F435W-F606W)\approx$ 0.0--0.2. Typical errors in these colors are 0.1 mags. 

A succinct description of the colors along the semi-minor axes is more difficult, as the errors are larger. Both colors are strongly affected by the scattering strip which produces a peak of emission at 3'' in the NE, and 3'' and 3.''7 at the SW. Outside these regions, $\Delta (F435W-F814W)\approx$ 0.3--0.5 mags, with indications of a decrease beyond 5'' along the SW side. 

The SBPs sample just narrow strips of all the data, and specially in regards to the colors may not give the full picture. Figures \ref{blue_colors} and \ref{colors} show the $\Delta(F435W-F606W)$ and $\Delta(F435W-F814W)$ colors of the disk relative to the star. We show the colors both before and after deconvolution. In both figures, the data have been convolved with a 4-pixel FWHM gaussian. Formally, the deconvolved image is the correct one to explore color information, but it is marred by streaks resulting from incorrect PSF matching, amplified by the Lucy-Richardson procedure. The effects of the scattering strip are manifested as two clumps both at the NE and SW sides of the star. As we mentioned before, the $\Delta(F435W-F814W)$ color is $\sim$0.12 mags less beyond 2.''5 in the deconvolved data.

These color maps make evident that the star is surrounded by an elongated nebulosity 8'' in size (measured along the semi-major axes). Beyond this limit there is an abrupt change in the characteristics of the dust, as only a weak color signal is detected (however, nebulosity is detected beyond this limit). Within the mid-disk, $\Delta (F435W-F814W)$ increases from the $\lesssim$0.5 at the SW side, to $\gtrsim$0.5 at the NE side, but it is mostly positive. If we assume that the SW side is closest to the Earth, this suggests a decrease in the value of the scattering factor $g$ as a function of increasing wavelength.  For $\Delta (F435W-F606W)$ the situation is less clean-cut, although overall it is also true that the color index is larger along the NE than along the SW side. Comparing figures \ref{minor} and \ref{blue_colors} indicates that the NE semi-minor axis is oriented along a particularly blue direction.

In section \ref{spiral} we noted that the arms have more neutral colors than the neighboring material. The effect is clearly seen in figure \ref{colors}, on which the arms can be roughly traced by the darker color. Along the NW side of the disk, $\Delta (F435W-F814W) \sim0.1$ mags within 3'', on the region occupied by structure 1. In the case of structures 2 and 3, the disk becomes neutral at the center of the features. (Along the smoothed SBPs the effect is only evident along the broad structure 2. See figure \ref{minor}.) The space within the arms has a non-neutral color in $\Delta (F435W-F814W)$.

\section{Discussion \label{discussion}}

As we have shown, these multicolor observations reveal a wealth of features in the circumstellar environment of HD~100546. Any model of this system should be able to reproduce the following characteristics:

\begin{itemize}
\item The circumstellar environment can be divided in three regions, with different brightness and color characteristics: the inner disk (from 1.''6 to 3''), the mid-disk (from 3'' to 8'') and the extended envelope (beyond 8''). For all distances and at all angles, the disk is intrinsically brighter in the F814W band than in the other bands.

\item The inner disk is the region of the spiral arms. In addition to the well-known spiral structure, there are smaller arms never before reported. If we assume that the disk rotates counterclockwise the main arms trail the direction of disk rotation. Colors of the arms are more neutral than those of the surrounding material. The space between them becomes brighter as one goes to longer wavelengths, even after deconvolving by the coronagraphic PSF. The morphology does not change as a function of wavelength.

\item The mid-disk is brighter and redder to the SW than to the NE. If the larger brightness is interpreted as the result of scattering asymmetry, it implies that the SW side is oriented towards the Earth. A description with a single value scattering asymmetry parameter implies $g\sim0.15$, although there are local variations suggesting values as large as $g\sim0.2$  The mid-disk has $\Delta (F435W-F814W)\sim 0.5$ and $\Delta (F435W-F606W)\sim 0.1$ although strong local variations are present. Comparison with NICMOS observations indicates that $\Delta (F435W-F160W)=1.5$ at 2.''5. The SW-NE color asymmetry suggests that $g$ decreases with increasing wavelength. 

\item An abrupt change of color and brightness marks the end of the mid-disk at 8'' along the major axis. Farther out, we detect nebulosity to 12''--14''. Overall, the shape is roughly consistent with what is expected from a flattened envelope with the same inclination as the inner disk. 

\item Power-law fits to the SBPs produce indices steeper than $-3$ to 8''. This dependence has also been observed in NICMOS and STIS data.
\end{itemize}

A detailed model of this system is beyond the scope of this paper, but in the following sections we explore possible interpretations of these observations that may serve as a basis for such a model. We argue that the behavior of the SBPs with distance suggest that we are not detecting the optically thick disk responsible for the far-IR emission in the SED but only the optically thin envelope. The spiral arms are then structures in this envelope.  Overall, the explanation for the colors of the dust remain a mystery, but we show that they are similar to colors of Kuiper Belt objects in the solar system. This suggests the processes like cosmic ray irradiation may be important to understand the dust color in HD~100546.

\subsection{The Color of the Circumstellar Material}

Given the large fractional infrared excess ($L_{disk}/L_{*}=0.51$), we expect some part of the circumstellar material to be optically thick at visible wavelengths.  This is not inconsistent with the very low relative scattered fluxes, as the amount of light scattered by an optically thick disk is controlled by the disk flaring angle. Neglecting the wavelength dependence of the scattering phase function, one expects that a situation in which every light path reaching the observer scatters off an optically thick structure to result in gray scattering. This is because interstellar dust albedo is approximately constant with wavelength in the optical \citep{whi95}. Therefore, the colors of the circumstellar material, especially those of material within $\approx$3'', which has been modeled as a flared, thick disk, are somewhat unexpected. On the other hand, optically thin lines of sight should result in blue scattering, if the dust around HD~100546 has the same optical constants and size distribution (with dust particles ranging in size from tens of Angstroms to a few microns) of the diffuse ISM \citep{dra03}.

By analyzing the extinction as a function of distance for B-stars near the line of sight to HD~100546, \citet{mal98} concluded that the reddening of the latter is consistent with the ISM column density at 100 pc. This means that the surrounding nebulosity does not contribute significantly to the reddening and the star is in front of the observed structures (or that the optical depth of circumstellar material in front of the star is $\tau_{\rm V} \lesssim 0.05$, from the error of the extinction measurements). In principle, some of the nebulosity could be associated with the nearby Lynds dark cloud DC~296.2-7.9, whose nominal center is only $\sim$25' away from the star. The ISM dust associated with the dark cloud should scatter more strongly in the blue than in the red. 

From figure \ref{major}, the brightness in F814W along the major disk axis would have to be 60\% smaller (or the brightness in F435W 60\% larger) for the scattering to be gray (using 0.5 mags as F435W-F814W color). This is much larger than the uncertainties in the measured stellar colors (Section \ref{stell_phot}). This red color is present even before the Lucy-Richardson deconvolution procedure, the effect of which is to reduce the F435W-F814W color at large angular distances. The disk is also red also in the F435W-F606W color, although not as strongly.

Multicolor, resolved,  scattered-light observations of disks around Herbig Ae/Be stars, that could be used to compare with our observations of HD~100546 are not very common. There are 14 Herbig Ae/Be stars for which the disk has been resolved\footnote{Catalog of Resolved Circumstellar Disks, C. McCabe, www.circumstellardisks.org/}. Of those, six or seven (in addition to HD~100546 these are: AB Aur, \citealt{gra99}; HD~163296, \citealt{gra00}; HD~150193A, \citealt{fuk03}; PDS144 N, \citealt{per06}; HD~142527, \citealt{fuk06}; and perhaps HD 169142, \citealt{hal06}) have been resolved in scattered light. Besides HD~100546, HD~142527 (F6, 2 Myr-old) and PDS144 N ($\sim$A2, age uncertain) have observations in multiple bands -- in the near-IR -- performed by the same group, ensuring that the photometric calibration is consistent. In the case of HD~142527 those observations reveal a gray disk. PDS144 N is an edge-on system, and so the disk appears as a dark gray lane. Observations of T Tauri stars are more common, and for the most part reveal gray disks (e.g. TW Hya, \citealt{wei02}). An interesting exception is GG Tau, which has a red, optically thick circum{\it binary} disk \citep{kri05a}. The large optical color of the disk relative to the star ($\Delta(V-I_{c})=0.8$, \citealt*{kri02}) has been attributed to a combination of reddened illumination by the central starlight upon passing through inner circumstellar disks and the scattering characteristics of the dust \citep{duc04}. Observations of edge-on T Tauri stars reveal outflow-cavity walls blue in color (relative to their illuminating stars), and whose optical characteristics can be explained by modified ISM extinction laws  \citep{sta03,wat04}.

If one considers the wavelength dependence of the phase function, it is possible to produce non-grey, optically thick disks. \citet*{mcc02} and \citet{duc04} argue that at least some fraction of the observed color of the GG Tau disk is due to a combination of the decrease of the scattering asymmetry parameter $g$ with increasing wavelength, and the inclination of the disk. If $g$ is large ($g\approx0.5$ in \citealt{mcc02}), this effect can produce very red disk colors relative to the star, even in optically thick disks. Figures \ref{blue_colors} and \ref{colors} show evidence that the phase function plays a role in the color of the disk, because the dust becomes redder as the scattering angle becomes larger (assuming that the SW side is closest to Earth). However, if the angular dependence of the scattering function can be modeled by a Henyey-Greenstein function, an explanation based solely on the wavelength dependence of the phase function does not reproduce the colors. The Henyey-Greenstein function is such that if backward scattering results in $\Delta(F435W-F814W)=0.5$, forward scattering should result in $\Delta(F435W-F814W)=-0.5$. In other words, if the back of the disk is very red, the front should be very blue. This is not what is observed in HD~100546. 

Here we discuss three additional possibilities to explain the color of the observed circumstellar material: (1) reddening due to dust very close to the star, (2) the dust characteristics of an optically thin envelope, and (3) selective absorption on the surface of the grains.

\subsubsection{Obscuring dust}

By analogy with GG Tau, we consider the possibility of obscuring material near the star which reddens the stellar flux illuminating the disk. Using the standard ISM extinction law with  a normalized optical extinction value $R_V$=3.1 \citep{car89}, we conclude that the 60\% excess in F814W with respect to F435W, requires A$_V=0.65$ mag to produce a gray disk. In other words, if there is a cloud of ISM material close to the star with  A$_V=0.65$ mag, the disk would appear red although it would really be gray. Structures 1, 2, and 3 would then be slightly blue. However, the SE Major axis F435W-F606W color is at most 0.2 mags, which implies A$_V=0.45$ mag. The same value of extinction can reproduce both colors if the extinction law changes more slowly in the optical. If $R_V\approx12$, the colors will match with A$_V\approx1.3$ mag. This would be in addition to the photometrically measured stellar extinction (A$_V$=0.28 mag). Such large value of $R_V$ has never been measured \citep{dra03}.

Some stellar obscuration is expected from the models developed by \citet{dul01}, due to shadowing of the starlight by the inner disk rim. For the particular case of HD~100546, such a rim is quite small, which suggests that this source of optical depth is not important at large distances from what we call the inner disk. Notice, however, that we detect red colors at all distances $\gtrsim$1.''6, well within the $\approx$4'' disk radius predicted by the models by \citet{dom03} mentioned in the introduction. 

\subsubsection{A Thin Envelope}

We may be detecting an optically thin envelope, as suggested by \citet{vin06}, made up of a mixture of particles whose total opacity increases with increasing wavelength.  This behavior of the opacity seems to be common of most debris disks with resolved circumstellar material in scattered light \citep{kri05}. For debris disk systems the circumstellar material is more evolved (there is less gas, dust grains are larger, the dust is optically thin) than in Herbig Ae/Be stars. Observations of HD~141569, a 10 Myr old A0 star surrounded by an optically thin debris disk show that the disk colors are $\Delta (F435W-F606W)=0.1$ and $\Delta (F435W-F814W)=0.25$ \citep{cla03} and recent observations of $\beta$ Pictoris reveal similar colors along the spine of its disk \citep{gol06}. No known debris disk has optical colors as large as HD~100546.

A scattering opacity increasing to the red will result if the longest wavelength of observation is comparable to the lower limit in the particle distribution, even for a standard astronomical silicate. The scattering efficiency is given by $Q=\sigma/\pi a^2$, where $\sigma$ is the scattering cross section and $a$ is the particle radius. Using the smoothed astronomical silicate from \citet{lao93}, and integrating over the bandpass (see Eqn. 8 of \citealt{gol06}) we obtain the effective scattering efficiency, $Q_{\rm eff}$ given the left panel of figure \ref{qeff}. The secondary peaks have been smoothed, as they are the result of the assumption of  perfectly spherical particles in the scattering opacities calculated by \citet{lao93}. 

$Q_{eff}$ has a broad maximum at a particle size comparable to the wavelength of observation divided by $4(n_\lambda-1)$, where $n_\lambda$ is the real part of the complex refraction index of the grain \citep{aug04}. Notice that the $Q_{\rm eff}$ from figure \ref{qeff} is not the same quantity derived by \citet{gol06}. In that work, the authors' ``effective scattering efficiency'' includes effects due to the dependence of the optical depth on the line of sight.

To compare with observations, one should integrate the effective scattering cross section over the range of particle sizes, weighted by the particle size distribution. The resulting observable, the normalized surface brightness, is given by:

$$ I/F_* \propto {\int_{a_{min}}^{a_{max}}}a^{2-p}Q_{eff}f_g(\theta) da$$

where $I$ is the observed surface brightness, $F_*$ is the observed stellar flux and $f_g(\theta)$ is the scattering phase function, which we assume to be a Henyey-Greenstein function. It depends on the particle size through $g$. We have assumed that the particle size distribution is $dn/da \propto a^{-p}$. The proportionality constant depends on the surface density of particles as a function of distance from the star.

The ratio of $I/F_*$ between two bands is shown in figure \ref{qeff} (right), where we plot the $\Delta (F435W-F606W)$, $\Delta (F435W-F814W)$ and the $\Delta (F435W-F160W)$ colors. For this figure we assume that $p=3.5$ \citep{doh69} and that the scattering angle is $\theta=$90 degrees. In other words, these should be the colors along the major disk axis.

If the lower limit in particle size is $\sim0.2$~$\mu$m, colors like $\Delta(F435W-F606W)\approx 0.2$ mags and $\Delta(F435W-F814W)\approx 0.5$ mags are possible, but the predicted $\Delta(F435W-F160W)$ color is too small, compared to the observations. If the lower limit in particle size is $\sim0.7$~$\mu$m, $\Delta(F435W-F160W)\approx 1.5$ mags, $\Delta(F435W-F606W)\approx 0.2$ mags, but $\Delta(F435W-F814W)$ is too large. Analogous results are obtained with pure graphite dust and with different grain porosities (as done in \citealt{gol06}). We also experimented with different particle size distributions, ranging from uniform distributions ($p=0$) to single-sized particles ($p=\infty$). The three colors cannot be reproduced simultaneously.

%The observed color depends not only on the lower limit in the particle size, but also on the shape of the size distribution. Figure \ref{qeff} (right) shows the result of the calculation assuming that all particles are the same size (i.e. that the particle size distribution is a delta function). With this assumption, we obtain that the maximum predicted colors are $\Delta(F435W-F606W) \approx 0.2$ and  $\Delta(F435W-F814W)\approx 0.5$ for particles with sizes $\approx 0.3-0.4 \mu$m. This is roughly consistent with observations. 

These arguments seek to explain the color of the scattered light as being a function of particle size, and assume that there is no selective absorption due to surface effects. The fact that the disk is not gray then implies the existence of ``small'' particles. For the dust model shown in the right panel of figure \ref{qeff}, the disk becomes neutral in the optical if the minimum size particles are $\gtrsim$5~$\mu$m. If the system does not have any gas, the lower particle size limit is set by radiation pressure. For HD~100546, $a_{min}$ is between $\sim$5 and $\sim$50 $\mu$m, depending on the exact dust model used (\citealt{tak01}; \citealt{li03}). The presence of particles smaller than the radiation pressure limit is possible in systems with even modest amounts of gas \citep{ard05}. Indeed, any explanation of the richly-featured ISO spectrum near 10 $\mu$m \citep{mal98} requires the presence of particles $\lesssim 2 \mu$m in size close to the star. For example, \citet{van05} models the 10 $\mu$m feature with dust whose crystalline component is dominated by forsterite grains 0.1 $\mu$m in radius.  

While models of the mid-IR spectrum consider dust much closer to the star than the one described in our observations, \citet{bou03} argue that small forsterite grains ($\lesssim 1\mu$m) are present in the disk at all distances. They attribute this result to collisional stirring by an inner ($\sim10$ AU) planet which expels small particles off the disk plane. The particles are then pushed out by radiation pressure tempered by gas drag. The lack of information about the gas content of the disk makes it impossible to predict the dynamics in detail. Segregation of dust by size is also possible if dust grains have grown and some have settled in the disk mid-plane. Smaller dust particles will remain suspended at larger scale heights (e.g. \citealp*{tan05}).

\subsubsection{Selective Absorption}
If there is selective absorption (i.e. if there are strong albedo changes in the observed wavelength range, beyond what is expected from models based on ISM-like dust) the previous analysis will be incorrect. The scattering characteristics of the dust may then be dominated by the optical constants of materials deposited on the surfaces of the grains.  In the solar system, Kuiper Belt and Centaur objects (KBOs) present relative colors with respect to the sun ranging from neutral to $\Delta(B-I)\approx2.5$ mags \citep{luu96}. This trend continues into the near-IR \citep{dels04}. Optical colors like those observed in HD~100546 are not uncommon (figure \ref{kbo}). In this context, the F160W observation is within the range of colors sampled by the KBOs, having a normalized reflectivity (to $V$) of $\approx 3$. 

It has been argued (see \citealt{luu96, dels04} and references therein) that the diversity of colors among KBOs and Centaurs is the result of the competing processes of formation of organic compounds on the icy surfaces (due to space weathering) and resurfacing (due to mutual collisions and cometary activity). The sources of space weathering are UV irradiation from the sun (most important in the case of solar flares), solar cosmic rays, and Galactic cosmic rays \citep{gil02}. Independently of whether or not this is the correct explanation for the observations presented here, weathering and resurfacing processes will also be at work around HD~100546, because the water-ice sublimation boundary (T$\sim$120 K) occurs at $R\approx30$ AU for large particles. 

\citet{dels04} developed a model of the reddening process in KBOs, in which the reddening timescale ($\tau\sim10^{8\pm1}$ yrs, \citealt{shu72}) is controlled by the Galactic cosmic ray flux. In the case of $\beta$ Pic, \citet{gol06} used this model (with the timescale scaled by the stellar luminosity) to conclude that the dust grains escape the system before they are significantly altered. However, this scaled model may not be applicable to $\beta$ Pic or HD~100546. As they are early type stars, their  astrosphere will be smaller than the sun's (reducing the role of stellar cosmic rays) and the effect of galactic cosmic rays will be more important. Furthermore, for HD~100546 the UV flux ($\lambda<1500$ \AA) at 300 AU is $\sim$300 times larger than the solar UV flux at 40 AU. Recent analysis of the colors of the debris disk around HR~4796 (an A0 star, $\approx10 $Myrs old) also suggest a scattering opacity similar to that of some KBOs (A. Weinberger, 2006, personal communication). In summary, space weathering is likely to play a role in our understanding of the colors of dust around A stars. However, a model appropriate for the conditions around those stars has not been developed.

\subsubsection{Summary}

Our observations reveal two separate, independent facts. First, the circumstellar material has a non-zero color relative to the star. This is true even within 3'', where the thermal emission has been modeled as being due to an optically thick disk.  The color suggests that there is scattering asymmetry in the dust, extinction close to the star, optically thin lines of sight, or selective absorption. Second, the circumstellar material is red, even to large distances of the star, suggesting that all has been processed in a similar way. As we have shown, extinction requires an anomalously large value of R$_V$ to explain both sets of colors: values of R$_V$ this large have not been observed. An optically thin disk halo composed of astronomical silicate cannot explain the measured scattering asymmetry. The colors are within the range of colors observed in KBOs, which are believed to be produced by selective absorption on the surfaces of dust gains.

As in GG Tau, it is possible that more than one process is at work in the system. Beyond 3'', the role on inner disk extinction on the light scattered off the mid-disk should be relevant for our understanding of the system. The intense UV radiation coupled with the effect of Galactic cosmic rays may redden the dust before is dissipated. These ideas await better modeling efforts.

Except for an explanation of the colors based only on extinction, the remaining explanations imply that the dust opacity increases with wavelength. The explanations based on optically thin scattering of a yet-to-be-determined dust mixture and that of a KBO-like material are not necessarily different, as the polymerized ices believed to be responsible for the KBO colors may provide the required dust characteristics. Formally, dust coated with this kind of ice mixture will be red, no matter what its optical depth. However, as we show below, there are other pieces of evidence that suggest that we are observing an optically thin dust distribution.

\subsection{The Envelope}
\label{geom}

Our fit to the SBPs within 2.''7 using the \citet{gra01} ranges (see bottom portion of table \ref{power}) is consistent with a r$^{-3}$ dependence, which suggests that the structure responsible for the scattered light is a geometrically thin, optically thick disk \citep{whi92}. This same r$^{-3}$ dependence has been observed by \citet{gra01} (from 1 to 2.7'') and \citet{aug01} (from 0.5 to 2.5''). However, the SBP for a flat disk is only $\approx 10^{-7}/\phi^2$ arcsec$^{-2}$, where $\phi$ is the field distance (in arcseconds) to the star \citep{whi92}. This is about 3 orders of magnitude less than what is observed. A flared, optically thick disk would result in a shallower SBP.  Notice that for AB Aurigae the H-band scattered-light surface brightness profiles decrease with stellar distance as r$^{-3}$ \citep{fuk04}. This near-IR dependence is steeper than the r$^{-2}$ dependence found by \citet{gra99} in the optical.

On the other hand, the SBP of an optically thin disk may present any radial power-law, depending on the surface density of the scatterers and/or changes in the scattering cross section as a function of distance. If the material is optically thin along the line of sight, the surface brightness profile then implies a surface density of r$^{-1}$ or r$^{-1.8}$ (depending on whether one uses the \citet{gra01} tracks or our 0.''25 wide tracks), ignoring variations in the scattering cross section. However, it is difficult to understand how the material around HD~100546 can be optically thin within 3'', because the dust in this region is the bulk source of the IR emission. A similar power law dependence is observed from 0.''5 to 3'' suggesting that the disk does not change from optically thick to optically thin within this range. It is unlikely that the transition to optically thin occurs within 0.''5 (50 AU) from the star as in this case, the spectral signature from the disk interior would appear much hotter than the $\approx50$K observed (e.g. \citealt{bou03}).  

An argument analogous to this led \citet{vin06} to posit the existence of a flattened optically thin envelope or halo surrounding a geometrically thin, optically thick disk. The disk surface (the surface where the optical depth from the observer to the disk $\approx1$) has to be shallow enough that it does not intercept significant amounts of the stellar radiation, but otherwise is unconstrained. The objective of  \citet{vin06} was to explain the 3 $\mu$m bump  mentioned in the introduction. They concluded that the envelope has constant optical depth in the visible given by $\tau_V=0.35$, although this refers only to relatively hot material responsible for the bump. In the context of this picture, our observations would only be detecting light scattered by the optically thin component. The optically thick disk is undetected in these optical observations. A remarkably similar picture has also been suggested by \citet{mee01} (see their figure 8, top panel) to explain the mid-IR SED. 
 
In summary, the distance dependence of the SBPs suggest the presence of an optically thick shallow disk with a geometrically thick, optically thin, flattened envelope, similar to that advanced by \citet{mee01} and \citet{vin06}. No thermal or scattering model with these characteristics has been published and so it is not clear if this picture may reproduce the SED and the large fractional infrared excess. A flat, optically thick disk can only produce  $L_{IR}/L_*=0.25$ \citep{ada86,whi92}. A more realistic, flared, optically thick disk, would need to have a total opening angle of $60^\circ$ to explain the excess. If the disk is flatter than this, the envelope would have to cover a larger angle (as seen from the star) and/or have a large radial optical depth. Our observations show an envelope extending to 1200 AU from the star, but whether this will be enough to reproduce the SED remains uncertain. 

%Calculations by \citet{aug01} indicate that the perpendicular (to the disk plane) optical depth is $\approx0.01$ at 2''. They conclude that is likely an underestimation, given that their calculated radial (parallel to the disk) optical depth is close to one.

\subsection{Spiral Arms or Shadows?\label{feat}}

Various explanations have been advanced to understand the spiral structure seen in the scattered-light images. On the one hand, it is possible that they are ``true'' arms, due to spiral density waves or tidal effects. \citet{qui05} conclude that the disk would have to be unreasonably thin for the Toomre {\it Q} parameter to be less than one, suggesting that the disk is stable to spiral-density waves. 

\citet{qui05} also examined the possibility that the spiral structures are driven by a compact object external to the disk, either a bound planet or a passing star. They performed hydrodynamical simulations (i.e. a pure gas disk) of the effect of bound planets and unbound stellar encounters on a 300 AU disk and concluded that a 20 M$_J$ mass planet/brown dwarf in an eccentric orbit would be necessary to excite the observed spiral structure, which they argued would be detectable in the NICMOS images. They also concluded that the hypothesis of an encounter with a passing star lacks suitable candidates. 

Could we detect such a body in our images? Figure \ref{detlim} shows the 3$\sigma$ detection limits for point sources around HD~100546. Because the differences in detection along the major and minor axes are small, we average them. To obtain these curves the number of counts has been transformed into instrumental vegamag magnitudes using the quantities from \citet{sir05}. Such transformation ignores color terms. Let us consider a 10 M$_J$ mass planet, 10 Myrs old. According to the models by \citet{bar03}, such an object has M$_{I_C}=14.8$, which implies m$_{I_C}=19.8$ at 100 pc, and m$_{\rm F814W}\approx19.4$, in the instrumental vegamag system. The V-I color of such an object would be 5.5 mags \citep{bar03}, and so it would be undetected in the F606W band. We searched for objects that were present in the F814W images but not in the F606W ones, but did not find any. Clearly, we would have been able to detect such object in any place outside the mask. This planet would have m$_{H}\approx16.3$ mags. \citet{aug01} do not provide detection limits as a function of distance for their observations, and so it is difficult to say if it would be visible in the NICMOS images. The objects predicted by \citet{qui05} to generate the structures would certainly be detectable in the images presented here. 

A number of researchers (e.g. \citealt{bou03, gra05, ack06}) have indicated that the presence of a hole in the gas and dust distribution at $\approx$ 10 AU suggests the presence of a planet. In this case, the perturber is internal to the disk, unlike the simulation by \citet{qui05}. The effects of this configuration on HD~100546 have not been explored.

Warps in the cast shadows on the outer disk that may appear as spiral arms. \citet{qui06} modeled the effect of warps on a disk that is optically thin perpendicular to the disk plane (within 80 AU) but partially optically thick along it.  In this model, the dark lanes are due to opacity of the mid-plane and the bright lanes represent optically thin material partially oriented along the line of sight. The model predicts that, as the disk is observed at wavelengths in which the opacity is larger, the dark lanes will change little, or they may become darker (because the mid-plane optical depth will increase) and the bright lanes will be brighter (as the optical depth along the line of sight increases). 

If we assume that the dust opacity increases from F435W to F814W, the brightening of the inter-arm space contradicts the models from \citet{qui06} and is more in tune with what one expects if the structures are ``real arms'': changes in the volumetric density of material separated from the inner disk by relatively empty regions, that become brighter at optically thicker wavelengths. However, as mentioned in section \ref{geom}, we believe the spiral arms are observed in the envelope, not in the optically thick disk. They may well reflect structures in the optically thick disk, but the latter is undetected.

In summary, the  observations presented here are inconsistent with the published simulations that seek to explain the origin of the spiral arms. Whether a model can be developed that generates the observed structure with an interior planet, with a very small exterior planet, or via warps in a more realistic, larger disk, remains uncertain.

%If the light from the star is being blocked by an ISM-like distribution of matter, and what we are seeing is a gray redden disk in which the opacity decreases from F435W to F814W, the inter-arm space does not change brightness between bands, while the structures become fainter at longer wavelengths. This is consistent with the models from \citet{qui06}. 

\section{Conclusions}

We present ACS/HRC coronagraphic images of HD~100546's circumstellar environment in the F435W, F606W, and F814W bands. The star is a B9.5, $\gtrsim$10 Myr old, 103 pc away from the sun. The observations were performed with the 0.''9 occulting spot. For each band, we observed the system in two different telescope rolls. To improve the contrast between the star and the circumstellar material, we subtracted a PSF reference star from HD~100546. We used HD~129433, which has a similar color as the target. To scale the two, we used direct measurements in the F435W band, coupled with color measurements of HD~100546 taken from the literature. We deconvolved the images using a simulated stellar PSF, obtained with the PSF simulator Tiny Tim. Both the deconvolved and non-deconvolved images in each band are available from the Journal Web page.

Subtraction residuals dominate the dataset within $\sim$1.''6. Color-combined images reveal a large scale structure of scattered light to distances $\approx$14'' away from the star. The well-known ``spiral arms'' (which we call structures 1, 2, and 3) of the disk are clearly seen. Photometric analyses by other groups reveal that the star is in front to the observed nebulosity. 

The scattered-light relative fluxes measured from 1.''6 to 10'' in the F435W, F606W, and F814W bands are $(L_{circum}/L_{*})_{\lambda}=1.3\pm0.2\ \times 10^{-3}$, $1.5\pm0.2\ \times 10^{-3}$, and $2.2\pm0.2\ \times 10^{-3}$, respectively. If we interpret the circumstellar nebulosity within 2'' as due to an inclined disk, the PA of the semi-major axis is $\approx 145$ degrees, with an inclination of $\approx$ 42 degrees. 

We analyzed the scattered light by taking median cuts along segments 0.''25 wide, oriented along the disk's principal axes. For these profiles, the 3$\sigma$ detection limits are $2\times 10^{-7}/arcsec^2$ (corresponding to  V$_{\rm Vega}$=23.4 magnitudes/arcsec$^2$). Analysis of these profiles reveal that the circumstellar material scatters more strongly towards larger wavelengths. Along the major axis, the $\Delta (F435W-F606W)$ color ranges from 0 to $\approx0.2$. The $\Delta (F435W-F814W)$ varies between $\approx0.5$  and $\approx$1, increasing slowly with distance.  Comparison with NICMOS results show that $\Delta (F435W-F160W)\approx$1.5. The SW side of the nebulosity is brighter (at all distances) than the NE side, by factors between 1 and 3, which we interpret as being due to forward scattering. This suggests that the SW side is the side closest to the Earth. The brightness ratio can be explained invoking a Henyey-Greenstein scattering phase function with $g=0.15$ (assuming an optically thin disk). However, there are local variations that suggests values of $g$ as large as 0.23. Evidence of a change in $g$ with wavelength is deduced by the fact the the SW side of the disk is bluer than the NE side.

Conceptually, we divide the disk in three regions, described by their size as measured along the major axis. The inner disk (from 1.''6 to 3'') has a size comparable to the optically thick disk used to explain the system's far-IR SED \citep{dom03}. This is also the region occupied by the spiral arms. The outer edge of the mid-disk (from 3'' to 8'') is marked by a factor of two decrease in brightness and by the disk color becoming almost neutral. Beyond that we detect nebulosity to $\approx$14'' along the major axis. The SBPs are consistent with the mid-disk and the extended nebulosity having the same inclination as the inner disk. 

The well-known spiral arms are clearly detected in the images. They are seen as a narrow bands of nebulosity that wrap around the disk. If the SW side is closest to Earth, spectroscopic observations that spiral structure trails the rotation. There are no significant morphological differences among the images in the different bands. The space between the arms and the part of the inner disk closer to the star becomes brighter at longer wavelengths, but the peak brightness of the structures does not change much, i.e. the structures become grayer at longer wavelengths. The brightening of the inter-arm space with wavelengths is inconsistent with models that assume that they are due to the effects of a warped disk. We deproject the disk and subtract a smoothed version of itself to increase the contrast of the sharp structures. This procedure suggest the existence of smaller, fainter arms, not easily seen in the original images. The NW arm along the major axis is split in two and the SW arm is revealed to have a companion along the NE side. The strongest arms seem to trace a four-armed spiral.

Dynamical models that seek to explain the structures as due to perturbations from external planets predict the presence of objects that are not seen in the images presented here. Observations of hot gas suggest the presence of a planet at 10 AU, but detailed models of the effect of this object in the disk of HD~100546 have not been published. The cause of these structures and the mechanism for their preservation remains unknown.

If one ignores the wavelength dependence of the scattering phase function, the fact that the inner disk has a color is surprising, given that the optical albedo derived for ISM grains is constant with wavelength, and that a large optical depth is predicted for this part of the disk. Most young  stellar objects in a similar evolutionary stage are gray. The more evolved debris disks have mostly red colors, although not as red as those of HD~100546.  If the colors were due purely to the behavior of the scattering phase function with wavelength, the disk would be bluer in the front side that what is observed. We examine the possibility that the color is due to the presence of dust between the star and the disk, beneath the coronagraphic mask. This would require different amounts of dust to explain different colors among the same direction, if the dust is ISM-like. $R_V$ would have to be $\approx12$ before we can reproduce the colors with the same amount of dust. We also explore the effect of the minimum particle size on the colors of a mixture of ``astronomical silicate'' grains. We conclude that a size distribution $dn/da \propto a^{-p}$ with $p=3.5$ is not capable of reproducing the colors. This general conclusion is also true for pure graphite grains, non-compact silicate grains, and for all values of $p$. In the context of an explanation for the color based on particle size, the color of the system implies the presence of particles $\lesssim$5$\mu$m at distances up to $\sim$800 AU from the star.

We show that the observed colors are within the range of colors observed in KBOs in the solar system. For KBOs the colors are believed to be the result of a competition between space weathering (cosmic ray and UV flux irradiation) and resurfacing due to collisions. This is a promising scenario and deserves to be explored more, by observing the circumstellar material at other near-IR bands. 

Within 2''.5, the major axis SBPs has a steep r$^{-3.8}$ dependence. This suggests that the structure responsible for the material is not an optically thick disk. A flat optically thick disk would result in SBPs with a r$^{-3}$ dependence on distance. A flared disk would result in a shallower power law. The distance dependence of SBPs requires invoking the presence of an optically thin halo or envelope in addition to the optically thick disk. Our observations are then consistent with the qualitative pictures from \citet{mee01} and \citet{vin06}, in which an optically thick disk is responsible for the far-IR emission but a large scale-height, optically thin disk (with a scale-height larger than predicted by standard disk models) surrounds the optically thick component. Given its colors, the envelope cannot be composed of remnant infalling ISM material but has to have been reprocessed by the star+disk system. The optically thick disk itself is undetected, suggesting that it has a very shallow surface profile. More detailed models are required before the characteristics of this envelope are better known.  

\begin{figure}
\epsscale{0.6}
\plotone{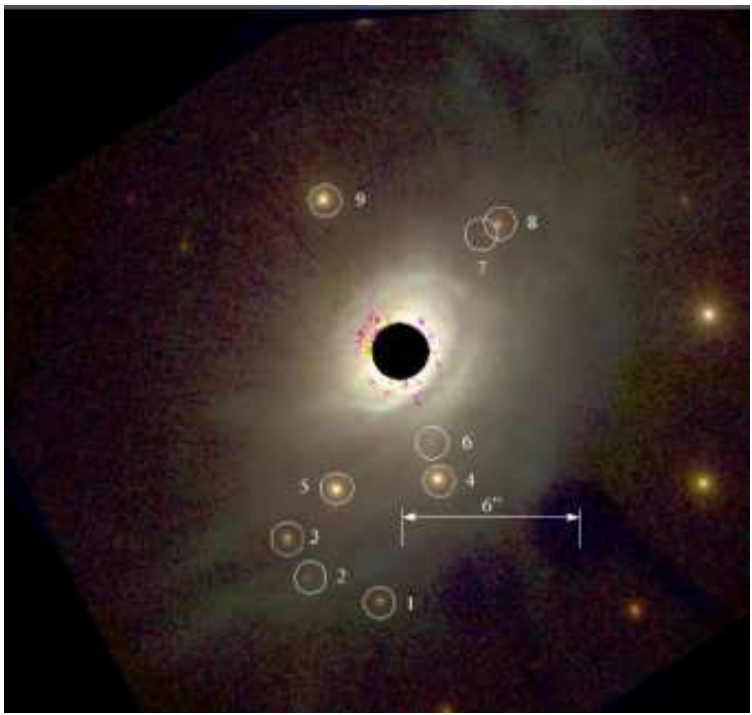}
\caption{\label{photo_stars} Guide to table \ref{field} (Color-combined image of non-deconvolved bands).}
\end{figure}

\begin{figure}
\epsscale{0.6}
\plotone{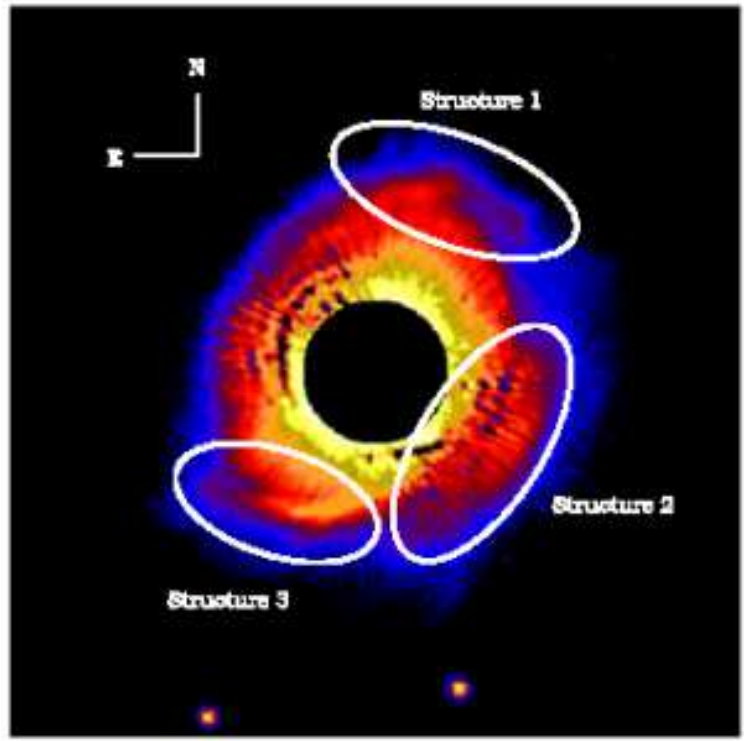}
\caption{\label{anot_struct}The inner circumstellar region as seen in the F435W band. We have annotated structures 1, 2, and 3, which are observed in all bands. The scaling is the same as in the mid-panels of figure \ref{panel1}. If the SW side is closes to Earth, spectroscopic observations suggest that the disk is rotating counterclockwise.}
\end{figure}

\begin{figure}
\epsscale{0.7}
\plotone{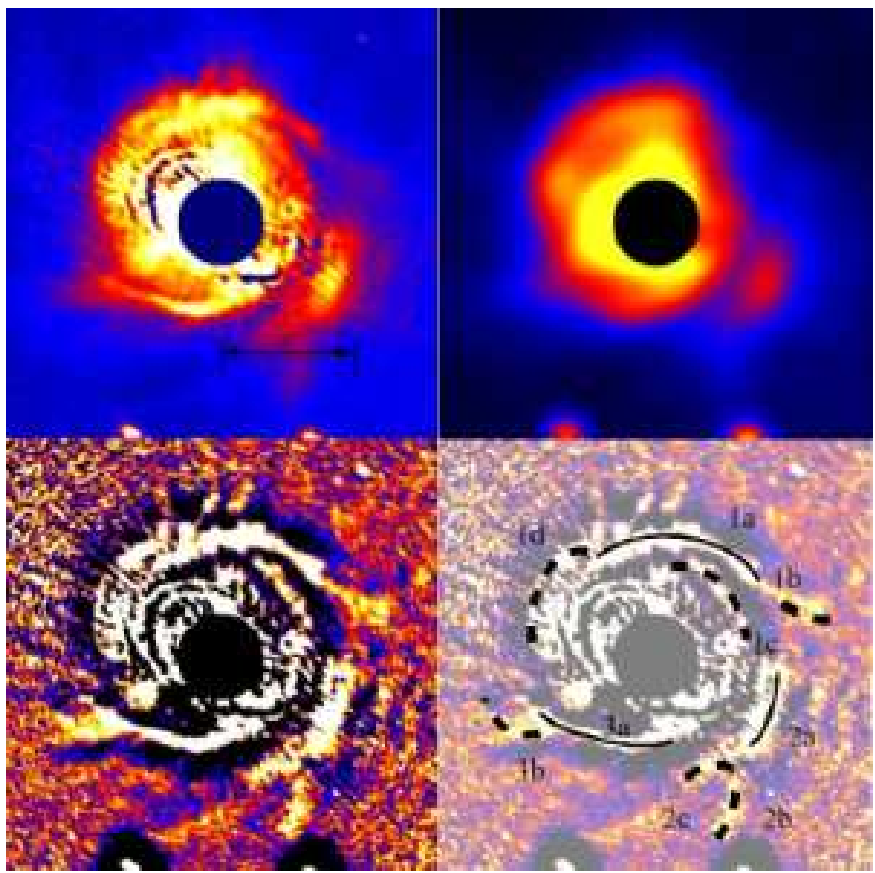}
\caption{\label{mosaic435} Unsharp masking of the non-deconvolved image in the F435W band. Upper left: Deprojected image, with each pixel corrected by scattering asymmetry ($g=0.15$), and multiplied by $\phi^2$, where $\phi$ is the angular distance to the star. Notice that the hemispheric subtraction residuals are circular in the original image, so they look elliptical here. Upper right: The same image, smoothed by a gaussian kernel with 30 pixels FWHM. Bottom left: Unsharp masking, the result of subtracting the upper-right image from the upper-left. Bottom right: Same as bottom left, with features identified. We only mark those features that are identified in all the bands. Features 1a, 2a, and 3a correspond to structures 1, 2, and 3 in figure \ref{anot_struct}.}
\end{figure}

\begin{figure}
\epsscale{1}
\plotone{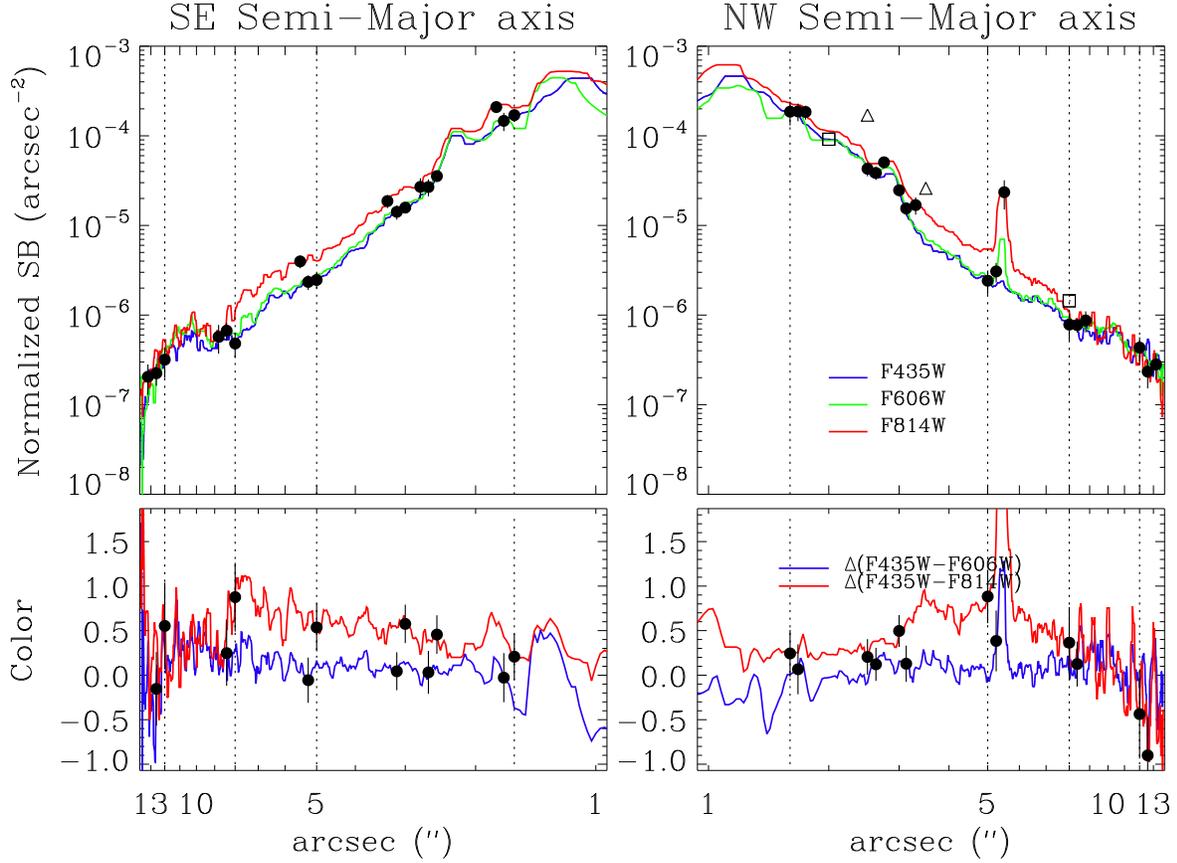}
\caption{\label{major}Top: Surface brightness profiles (SBPs) of the deconvolved images divided by the stellar flux. Bottom: Disk colors relative to the star. The SBPs are calculated by taking the median flux within segments 10 by 1 pixels (0.''25 by 0.''025) along an angle 145$^o$ East of North (the PA of the disk), to 13'' away from the star. In the plot, the profiles have been also smoothed using a median boxcar 10 pixels long. The dotted lines mark 1.''6, 5'', 8'', and 12'', corresponding to the circles shown in figure \ref{circles} and the ranges for the power-law fits (table \ref{power}). The errors are shown only at representative points, as filled circles. On the scale of the top panels they are of the order of the size of the circles. Subtraction residuals dominate within 1.''6. In the top right panel, the squares are STIS measurements \citep{gra01} and the triangles are NICMOS measurements in the F160W band \citep{aug01}.} %updated
\end{figure}

\begin{figure}
\epsscale{1}
\plotone{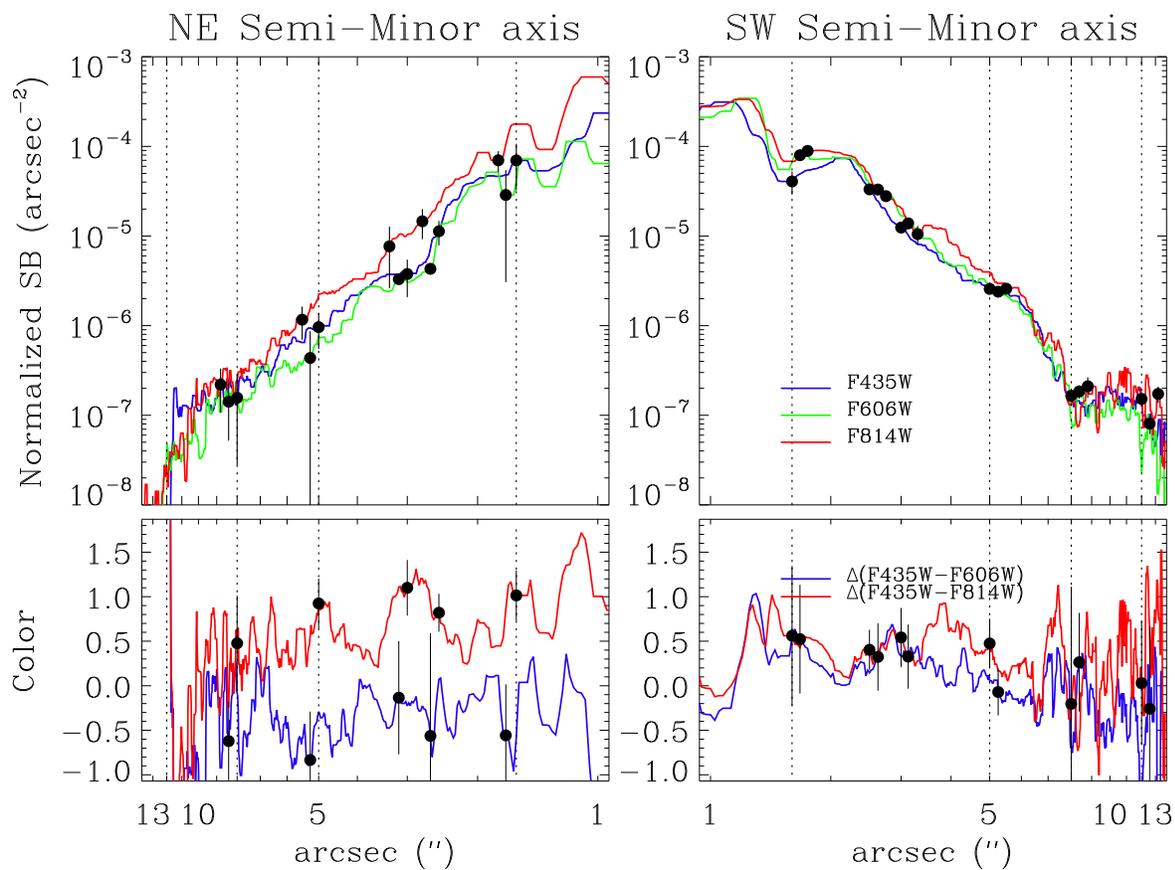}
\caption{\label{minor}Same as in figure \ref{major} but centered at an angle 55$^o$ East of North. Between 2.''5 and 3.''7 in the NE direction, we see the effect of the scattering strip as an increase in the colors. Beyond 11.''75 in the NE, we reach the end of the region sampled by both rolls in F435W. Beyond 8'' in the SW the profiles are affected by the large coronagraphic spots.}
\end{figure}

\begin{figure}
\plotone{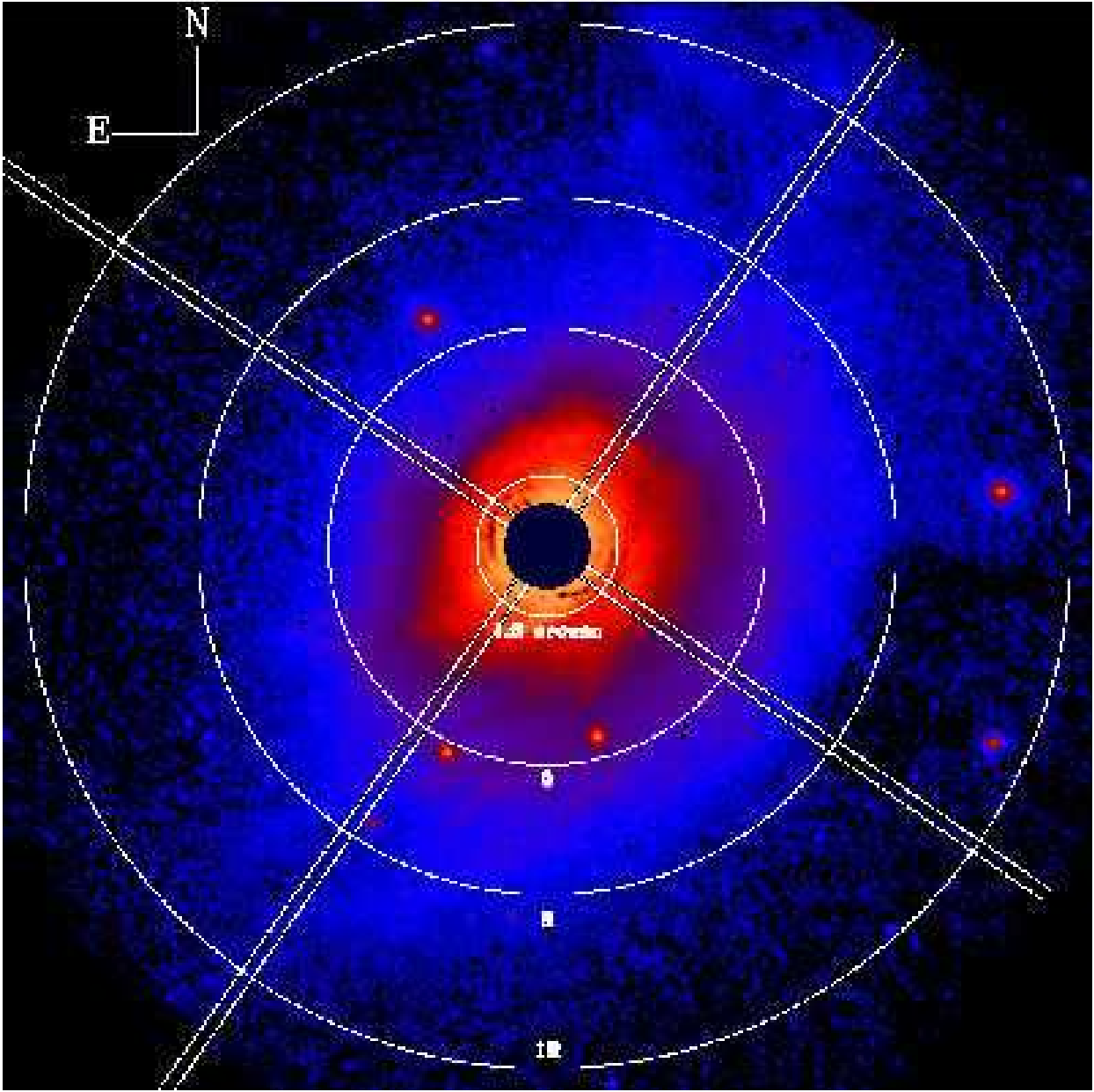}
\caption{\label{circles}Annotated, deconvolved image of F435W divided by the stellar flux, to use as a guide interpreting the surface brightness profiles of figures \ref{major} and \ref{minor}. The circles indicate radii 1.''6, 5'', 8'', and 12'', corresponding to the limits of the power-law fits. The rectangular boxes are 0.''25 wide, and are oriented in the position angles extracted to produce the SBPs. The image is in a logarithmic stretch, with limits $5\times10^{-6}$ arcsec$^{-2}$ to $10^{-3}$ arcsec$^{-2}$.}
\end{figure}

\begin{figure}
\plotone{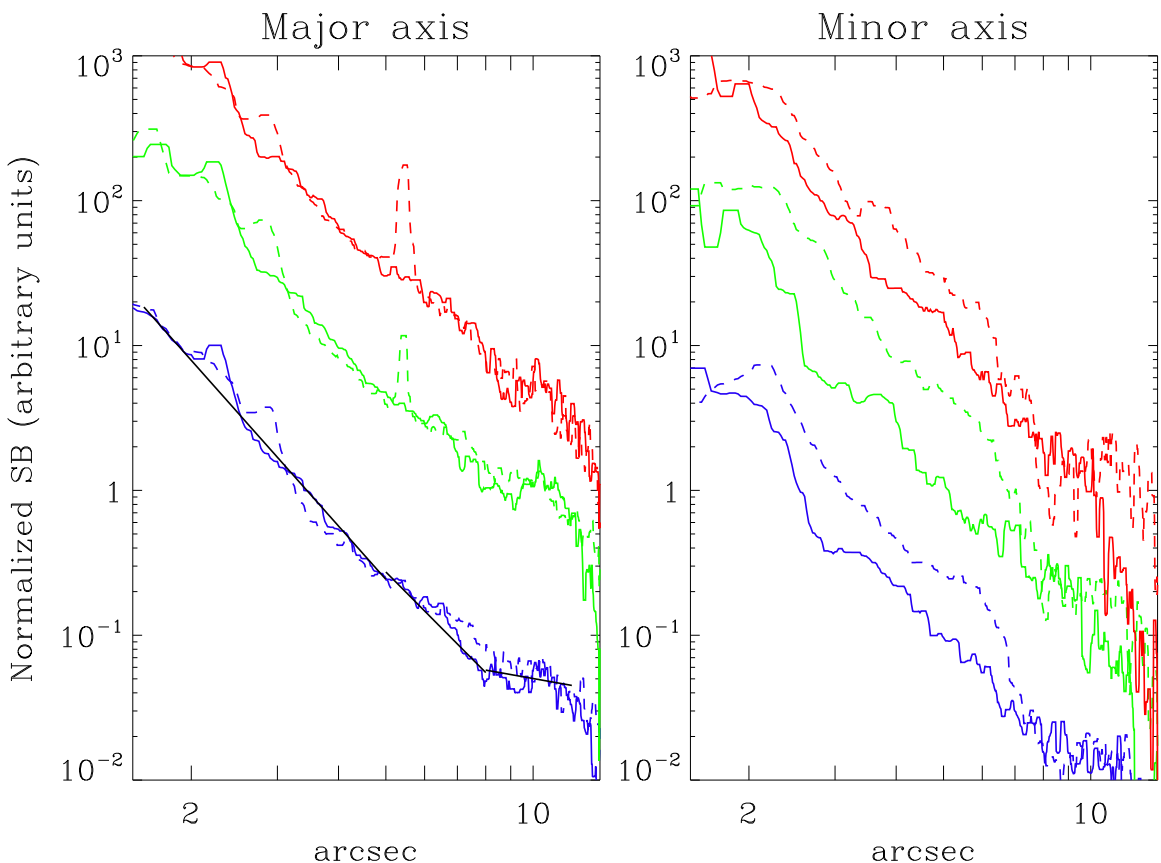}
\caption{\label{reflec}Comparison of the profiles at both sides of the disk. The profiles of different bands have been shifted by an arbitrary amount. Blue, green, and red colors correspond to F435W, F606W, and F814W respectively. (Left) Solid: SE side; Dashed: NW side. (Right) Solid: NE side; Dashed: SW side. Also shown are the power-law fits for the F435W band in the major axis (starting closest to the star, the indices are -3.8, -3.4, and -0.6)} %updated
\end{figure}

\begin{figure}
\epsscale{1}
\plottwo{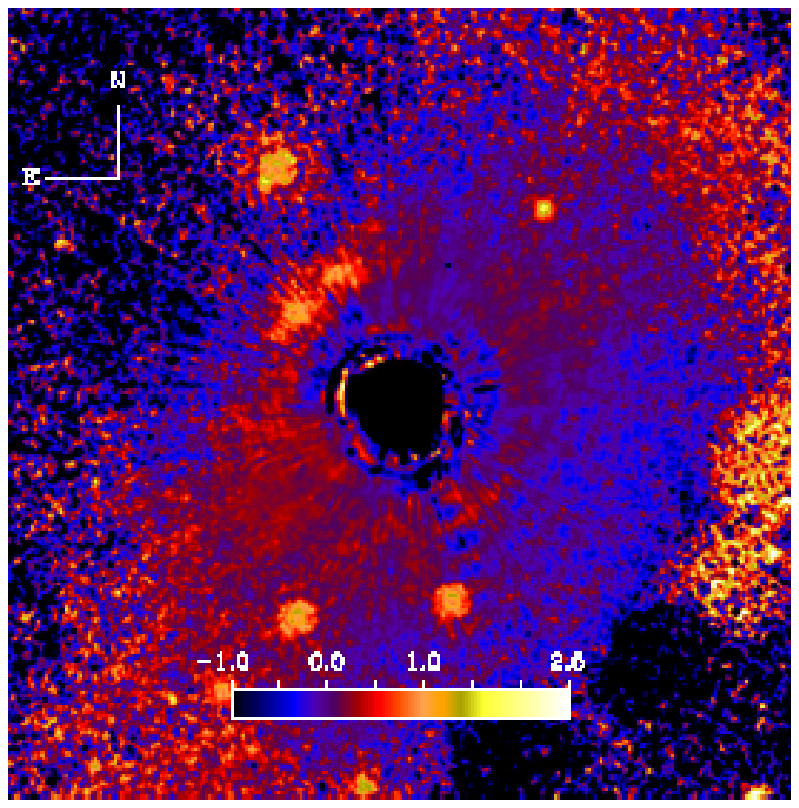}{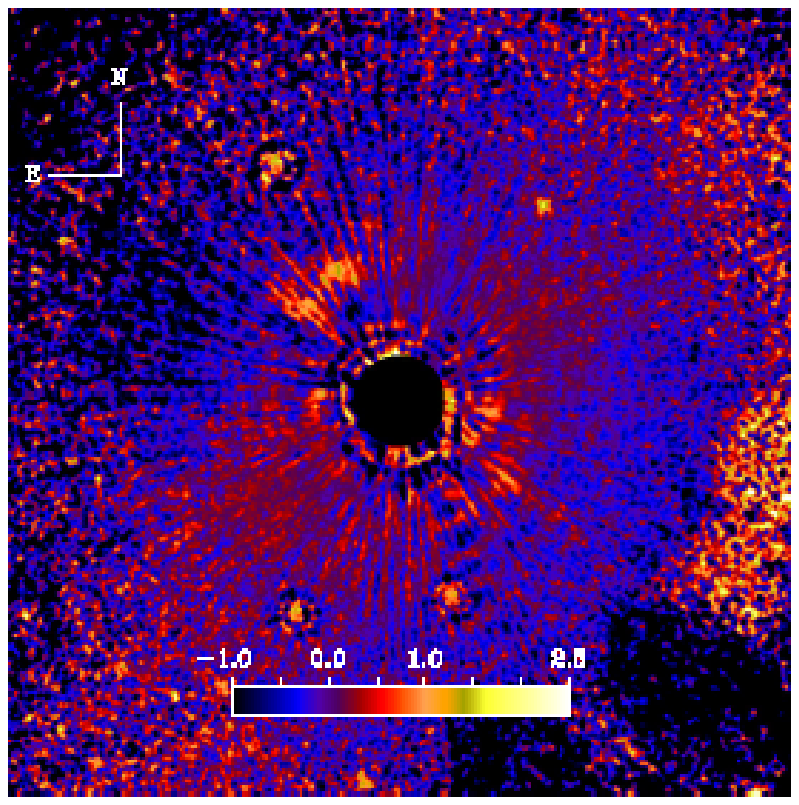}
\caption{\label{blue_colors} The $\Delta (F435W-F606W)=-2.5 \log (SBP_{F435W}/SBP_{F606W})$ color. Left: Color image obtained from non-deconvolved data. Right: Color image obtained from deconvolved data. In the NE the effect of the scattering strip is seen as a brightening of the colors. These features have weaker counterparts on the SW.} %updated
\end{figure}

\begin{figure}
\epsscale{1}
\plottwo{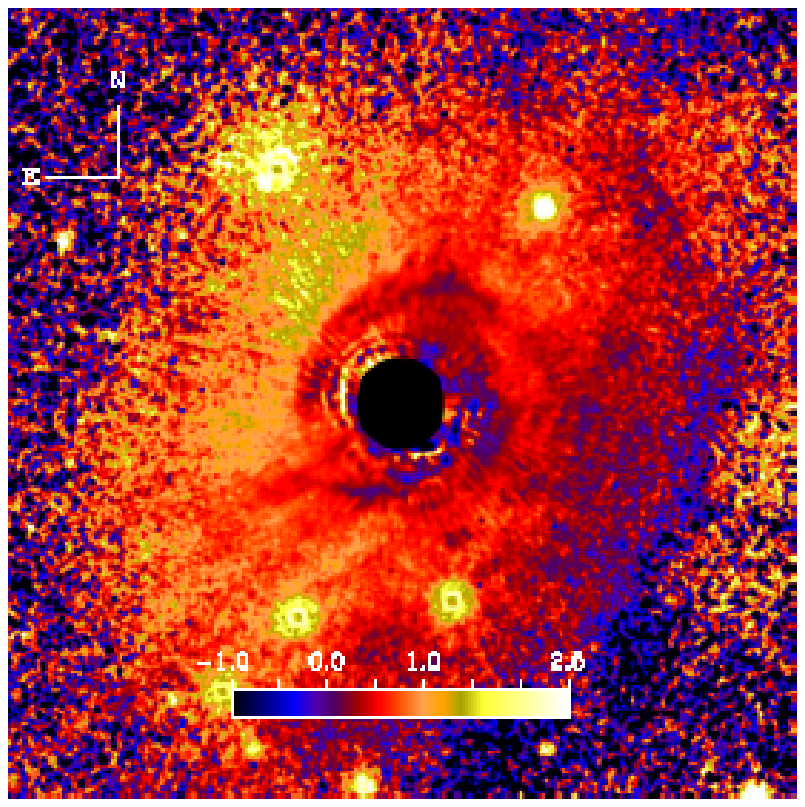}{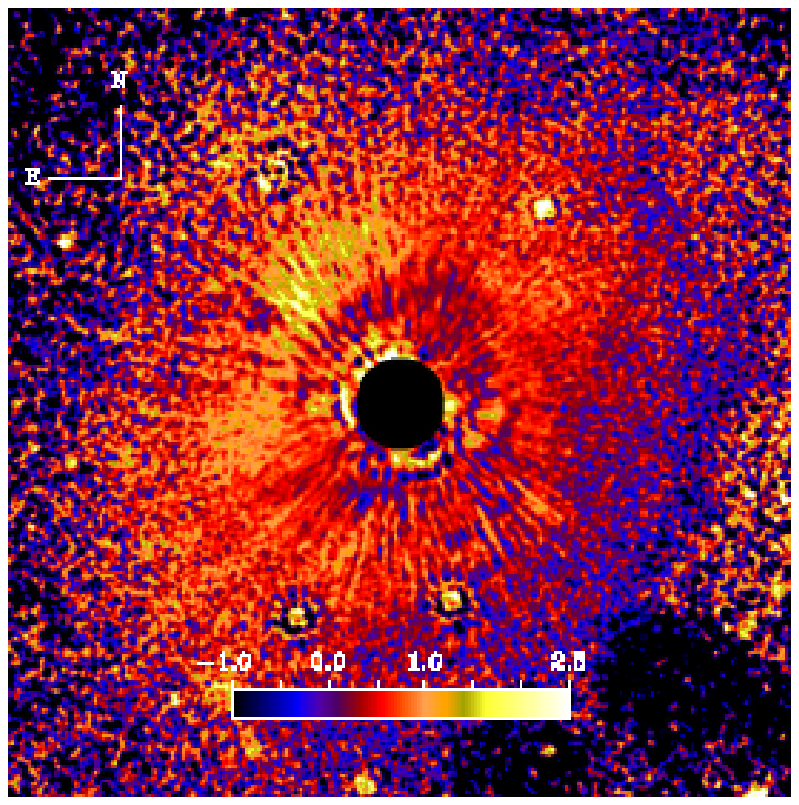}
\caption{\label{colors}Same as figure \ref{colors} but with $\Delta (F435W-F814W)=-2.5 \log (SBP_{F435W}/SBP_{F814W})$.} %updated
\end{figure}

\begin{figure}
\epsscale{1}
\plottwo{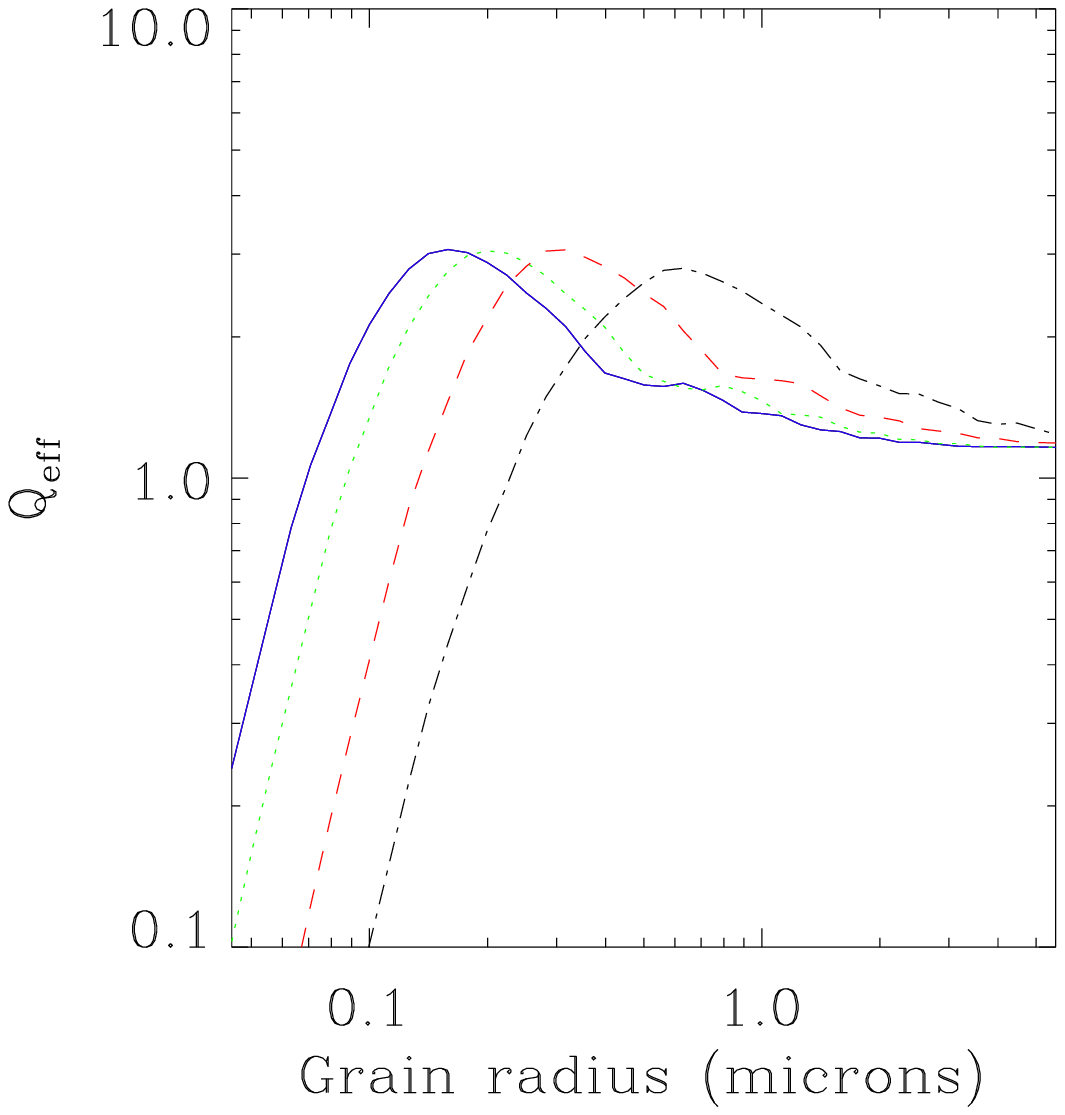}{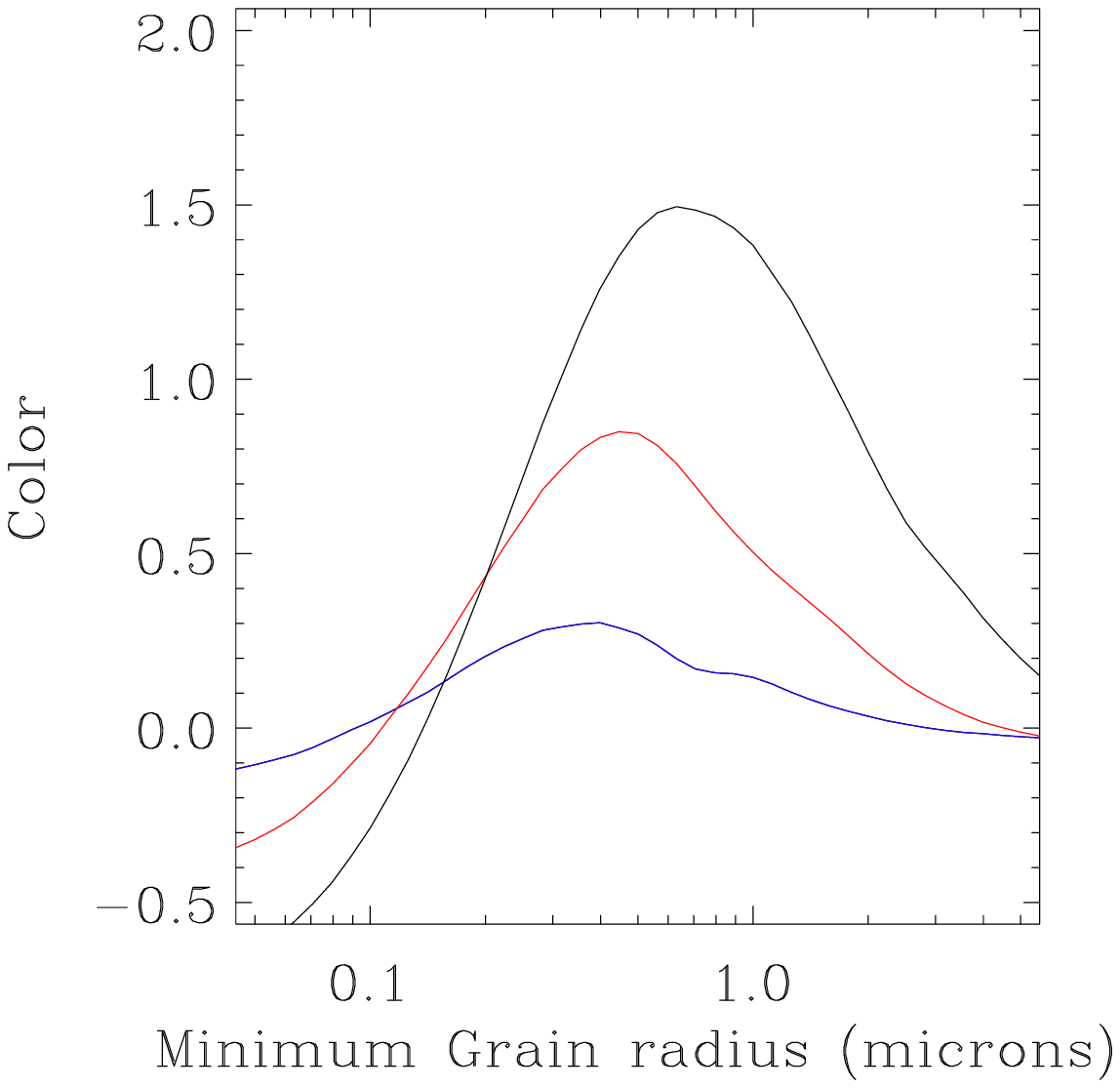}
\caption{\label{qeff}The color of the dust. Left: The scattering efficiency (averaged over observation band) as a function of grain size for the standard astronomical silicate. Blue (solid): F435W; green (dotted): F606W; red (dashed): F814W; black (dot-dash): F160W. Right: Expected color of the dust as a function of minimum particle size. We assume a size distribution in the form of a$^{-p}$, where p=3.5. Blue: $\Delta(F435W-F606W)$. Red: $\Delta(F435W-F814W)$. Black: $\Delta(F435W-F160W)$.}
\end{figure}

\begin{figure}
\plotone{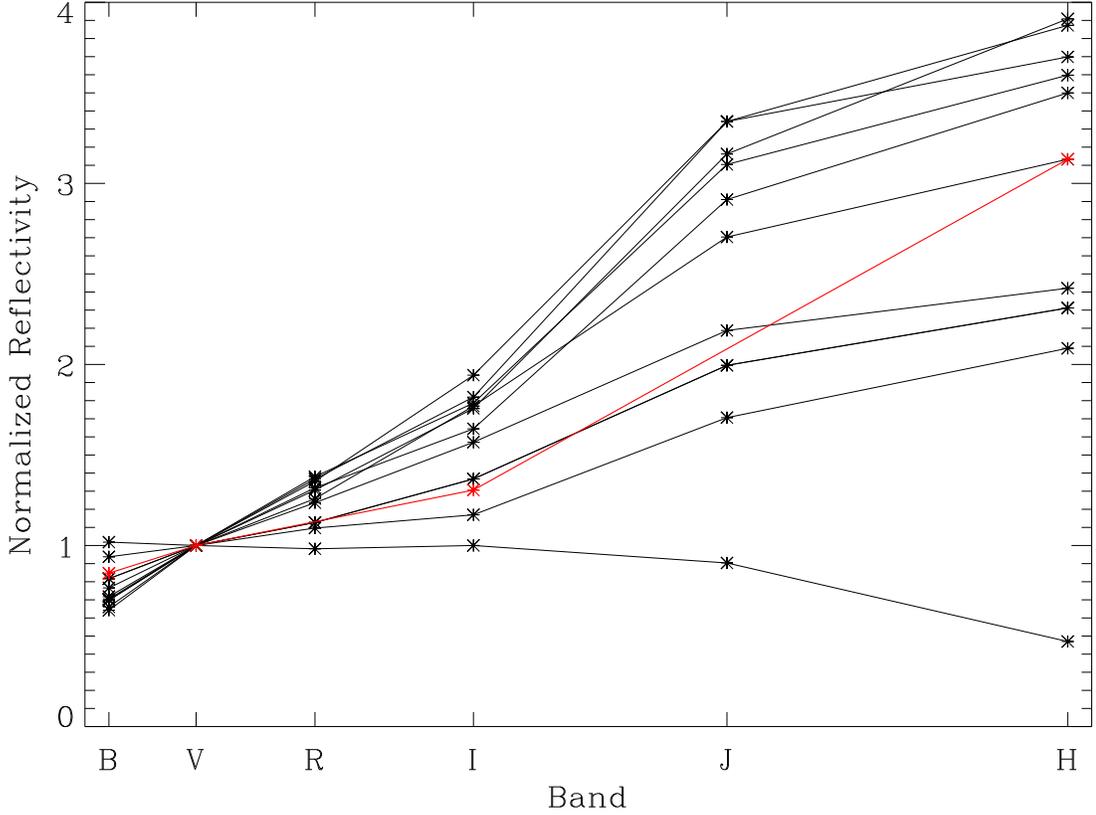}
\caption{\label{kbo}Colors of KBO and Centaurs (black lines), compared with the measured colors of HD~100546. The ordinate axis is the normalized reflectivity, $R_N=10^{0.4 \Delta(m_\lambda-V)}$, where $V$ is the star's magnitude \citep{luu96}. For the HD~100546 we use the colors measured 3.5'' away from the star along the SE semi-major axis, $\Delta (F606W-F435W)=-0.18$, $\Delta (F606W-F814W)=0.29$, and $\Delta (F606W-F160W)=1.24$. The colors were transformed from the instrumental to the standard vegamag system. The data for the solar system objects was taken from table 2 of \citet{dels04}. We include only those objects with BVRIJH photometry. Notice that the colors of the circumstellar material are well within the range of KBO colors.}
\end{figure}

\begin{figure}
\plotone{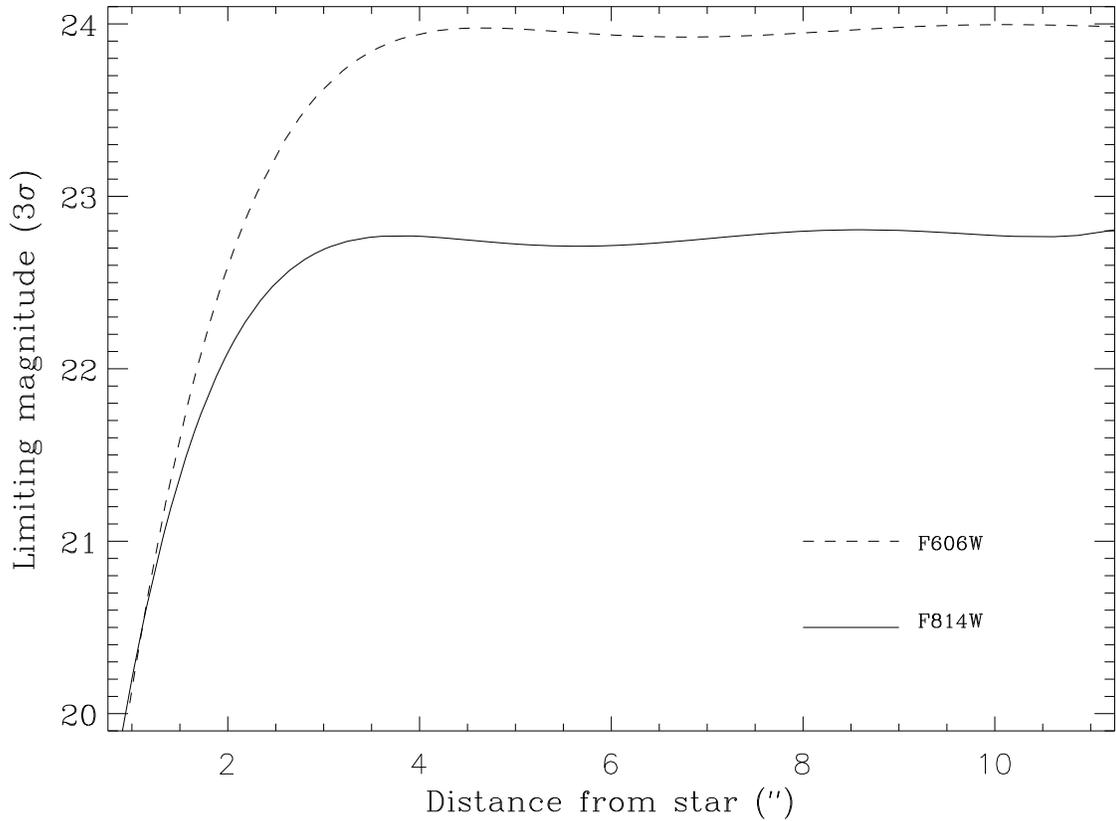}
\caption{\label{detlim}Detection limits for point sources. The curves show the 3$\sigma$ detection limits for point sources in F606W (dashed) and F814W (solid) magnitudes. The curves were obtained by planting stellar sources at different distances from the star and calculating how bright would those sources have to be to be detected at the 3$\sigma$ level. The noise is calculated based on the local background value. These detection limits are averages of detection limits along the same NE, SE, SW, and NW directions used to obtained the profiles from figures \ref{major} and \ref{minor}.}%The horizontal solid line shows the F814W magnitude expected for a 10 Myr old, 10 M$_J$ planet.
\end{figure}

The crowded field of HD~100546 suggests a way to probe the envelope material, by taking spectra of stars 3, 4, 5, 8, and 9 (Figure \ref{photo_stars}). However, for optically thin dust, the extinction in $V$ due to the envelope is of the order of $SB \times 4 \pi \phi^2$, where $SB$ is the normalized surface brightness at F606W, and $\phi$ is the angular distance between HD~100546 and the measurement position. This implies that at 5'', A$_V \approx 0.001$ mags. While this is a lower limit (some light may be blocked by the inner disk) trying to detect a small amount of extinction on the spectrum of a star 5'' away from a source forty thousand times brighter (comparing star 4 to HD~100546), is a formidable obstacle for this kind of experiment. 

Further observations and models are required before we can fully understand this system. In particular, it is crucial to measure the cool gas content of the circumstellar material, which will determine the dynamics of the dust \citep{ard05}. Additional coronagraphic observations at other NICMOS bands will provide much needed color points at longer wavelengths, which will help decide whether or not a new kind of dust model is needed to understand old protoplanetary (and/or young debris) disks. The observations presented here show how powerful multicolor coronagraphic data can be in advancing the understanding of protoplanetary disk evolution.

\acknowledgments
The authors wish to thank Mario Van den Ancker and Dejan Vinkovi\'{c} for their willingness to provide guidance regarding our interpretations of their work. Richard White was kind enough to provide us with the results of his PSF simulations in advance of publication, as well as extensive guidance on the use of the Lucy-Richardson algorithm. Dale Cruikshank was kind enough to illuminate the intricacies of KBO colors for us. Karl Stapefeldt provided advice in frequent discussions on the astrophysics of HD~100546. ACS was developed under NASA contract NAS 5-32865, and this research has been supported by NASA grant NAG5-7697. We are grateful for an equipment grant from Sun Microsystems, Inc. The Space Telescope Science Institute is operated by AURA, Inc., under NASA contract NAS5-26555. We are also grateful to K. Anderson, J. McCann, S. Busching, A. Framarini, and T. Allen for their invaluable contributions to the ACS project at JHU. This research has made use of the Catalog of Resolved Circumstellar Disks, the NASA's Astrophysics Data System Bibliographic Services, and of the SIMBAD and Vizier databases, operated at CDS, Strasbourg, France. 

%\bibliographystyle{/local/d1/bibtex/apj}
%\bibliography{/local/d1/bibtex/hd_disk}

\clearpage
\input{tab1}
\input{tab2}

\input{tab3}
\input{tab4}
\input{tab5}
\clearpage

 %updated

\end{document}

%% file: tab1.tex
%\documentclass{aastex}
%\begin{document}
\begin{deluxetable}{lccccc} 
\tabletypesize{\small}
\tablecolumns{6} 
\tablewidth{0pc} 
\tablecaption{Log of Coronagraphic Exposures\label{obs_log}}
\tablehead{
\colhead{Object} &
\colhead{Band} & 
\colhead{Obs. Start} & 
\colhead{Total Exp.Time} & 
\colhead{P.A.\tablenotemark{a}}&
\colhead{Gain} \\
	&	& (UT) & (secs)	& (deg.) &(e$^{-}$/photon)  }
\startdata 
HD~100546 & F435W & 2004-04-26 13:51:34 & 3   & -138.31 & 4 \\
          &       & 2004-04-26 22:39:00 & 160 & -138.35 & 2 \\
          &       & 2004-04-26 22:43:00 & 2400& -138.35 & 2 \\
          &       & 2004-04-26 23:54:00 & 160 & -118.35 & 2 \\
          &       & 2004-04-26 23:58:00 & 2530& -118.35 & 2 \\
          & F606W & 2003-03-26 19:42:39 & 3   & 179.09 & 2 \\
          &       & 2003-03-26 03:53:00 & 130 & 179.05 & 2  \\
          &       & 2003-03-26 03:57:00 & 2600& 179.05  & 2  \\
          &       & 2003-03-26 07:05:00 & 130 & -152.95 & 2  \\
          &       & 2003-03-26 07:09:00 & 2600 & -152.95 & 2 \\
          & F814W & 2003-03-26 19:42:39 & 6    & 179.09 & 2 \\
          &       & 2003-03-26 02:25:00 & 130  & 179.05 & 2 \\
          &       & 2003-03-26 02:28:00 & 2350 & 179.05 & 2 \\
          &       & 2003-03-26 05:33:00 & 160  & -152.95 & 2 \\
          &       & 2003-03-26 05:37:00 & 2520 & -152.95 & 2 \\
HD~129433\tablenotemark{b} & F435W & 2004-04-27 13:56:20 & 2 & 151.52 & 4 \\
          &       & 2004-04-27 01:32:00 & 16 & 151.48 & 2 \\
          &       & 2004-04-27 01:34:00 & 160  & 151.48& 2 \\
          &       & 2004-04-27 01:40:00 & 400  & 151.48& 2 \\
          &       & 2004-04-27 01:48:00 & 1300 & 151.48 & 2 \\
          & F606W & 2004-04-13 19:45:02 & 1  & 117.76 & 2 \\
          &       & 2003-03-26 09:33:00 & 16   & 117.72 & 2 \\
          &       & 2003-03-26 09:34:00 & 1020 & 117.72 & 2 \\
          & F814W & 2003-03-26 19:45:11 & 2    & 117.72 & 2 \\
          &       & 2003-03-26 09:13:00 & 16   & 117.72 & 2 \\

\enddata
\tablenotetext{a}{Position angle of image y axis (deg. E of N)}
\tablenotetext{b}{PSF Reference star for HD~100546}
\tablecomments{Only observations made with the parameter GAIN=4 preserve the number of counts in a saturated image.}
\end{deluxetable}
%\end{document}

%% file: tab2.tex
\clearpage
\begin{deluxetable}{rcc} 
\tablecolumns{3} 
\tablewidth{0pc} 
%\tabletypesize{\scriptsize}
\tablecaption{Photometry\label{photom}}
\tablehead{
\colhead{} & \colhead{HD~100546 (B9.5Vne)} & \colhead{HD~129433 (B9.5V)} }
\startdata 
			&F435W	   		&			\\
\cline{1-3} \\
e$^-$/sec ($10^7$)	&	2.5 (3\%)	&	6.4 (3\%)	\\
Flux Density (Jy)	&	8.3 (3\%)	&	21.2 (3\%)	\\
B$_{\rm Vega}$  \tablenotemark{a}	&	6.70 (0.03)	&	5.69 (0.03)	\\
\cline{1-3} \\
	   		&	F606W		&		\\
\cline{1-3} \\
e$^-$/sec ($10^7$)	&	4.9 (4\%)	&	11.1 (3\%)	\\
Flux Density (Jy)	&	7.3 (4\%)	&	16.4 (3\%)	\\
V$_{\rm Vega}$  \tablenotemark{a}	&	6.69 (0.04)	&	5.79 (0.03)	\\
\cline{1-3} \\
   			&	F814W		&		\\
\cline{1-3} \\
e$^-$/sec ($10^7$)	&	1.9 (5\%)	&	4.1 (3\%)	\\
Flux Density (Jy)	&	5.4 (5\%)	&	11.5 (3\%)	\\
I$_{\rm Vega}$ \tablenotemark{a}	&	6.64 (0.05)	&	5.82 (0.03)	\\

\enddata
\tablenotetext{a}{Vegamag system. The instrumental system produces the same magnitudes within the errors.} 
\tablecomments{HD~129433 is the star used a PSF reference. For F435W the count rate (e$^{-}$/sec) is measured from the image. The F606W and F814W count rates are estimated from the F435W values and synthetic photometry using proxy spectra. The F435W measurements have errors of $\sim$3\% \citep{sir05}. For HD~100546, the errors in F606W and F814W include errors in the measured colors of the target star in these bands, while the errors in HD 129433 are only those of the measured photometry in F435W.}   
\end{deluxetable}

%% file: tab3.tex
\clearpage
\begin{deluxetable}{rccc} 
\tablecolumns{4} 
\tablewidth{0pc} 
\tablecaption{Standard Photometry of Field Stars\label{field}}
\tablehead{
\colhead{Number} &\colhead{B} & \colhead{V} & \colhead{I} }
\startdata 
       1      & $23.7\pm0.5$  & $22.2\pm0.1$   &     $20.8\pm0.1$ \\
       2      & $>24.5$       & $24.2\pm0.3$        &     $22.1\pm0.3$  \\
       3      & $22.1\pm0.2$  & $ 21.2\pm0.09$  &     $20.2\pm0.1$ \\
       4      & $19.00\pm0.05$& $18.11\pm0.05$ &     $17.14\pm0.05$ \\
       5      & $19.37\pm0.06$& $18.42\pm0.03$ &     $17.44\pm0.05$ \\
       6      & $22.1\pm0.2$  & $21.8\pm0.1$   &     $21.1\pm0.2$ \\
       7      & $>24.0$       & $23.8\pm0.3$   &     $22.5\pm0.3$ \\
       8      & $23.2\pm0.4$  & $21.3\pm0.1$   &     $19.00\pm0.06$ \\
       9      & $19.5\pm0.06$  & $18.60\pm0.05$ &     $17.64\pm0.05$ \\

\enddata
\tablecomments{Aperture photometry for field stars, obtained from non-deconvolved images. See figure \ref{photo_stars} for a key to the numbers. Color transformations from the instrumental to the standard system given by \citet{sir05}. The error includes uncertainties in the measurement of the local sky and Poisson noise. It does not include the (systematic) errors introduced by transformation to the standard photometric system.}
\end{deluxetable}

%% file: tab4.tex
\clearpage
\begin{deluxetable}{lcc}
\tablecolumns{3} 
\tablewidth{0pc} 
\tablecaption{Geometric Parameters for the Disk\label{geometry}}
\tablehead{\colhead{Reference} &  \colhead{Inclination (deg from face-on)} & \colhead{PA of Major Axis}}

\startdata 
This Work & $42\pm5 \tablenotemark{a}$ & $145\pm5$ \\
Pantin et al. (2000) & $50\pm5$ &  $127\pm5$ \\
Grady et al.(2001) & $49\pm4$ & $127\pm5$ \\
Augereau et al. (2001) & $51\pm3$ & $161\pm5$\\
\enddata
\tablenotetext{a}{From isophot fitting between 1.''6 and 2'' from the star. The numbers quoted are averages for all bands. This determination ignores the angular dependence of dust scattering.}
\end{deluxetable}

%% file: tab5.tex
\clearpage
\begin{deluxetable}{cccc}
\tablecolumns{4} 
\tablewidth{0pc} 
\tabletypesize{\scriptsize}
\tablecaption{Power-law fits to the SE semi-major axis\label{power}}
\tablehead{\multicolumn{4}{c}{SE Major Axis} \\
\cline{1-4} \\
\colhead{Distance} & \colhead{F435W}   & \colhead{F606W}    & \colhead{F814W}} 

\startdata 
1.''6-5.''0&$-3.8\pm0.1$&$-3.8\pm0.1$ &	$-3.7\pm0.1$\\
5.''0-8.''0&$-3.4\pm0.3$&$-2.8\pm0.3$ &	$-2.7\pm0.3$\\	
8.''0-12.''0&$-0.6\pm0.4$&$-0.4\pm0.3$&$-1.2\pm0.5$\\

\cutinhead{\citet{gra01} ranges} 
1.''6-2.''7&$-3.6\pm0.4$ &$-3.5\pm0.4$ &$-3.5\pm0.3$\\
2.''7-5.''0&$-4.2\pm0.2$ &$-4.2\pm0.2$ &$-3.9\pm0.2$\\			
5.''0-8.''0&$-2.5\pm0.3$ &$-2.5\pm0.3$ &$-3.0\pm0.3$\\
	
\enddata
\tablecomments{The ``SE Major Axis'' fits are obtained along a strip 0.''25-wide, centered on the star and with PA=145 deg.  For the \citet{gra01} ranges, PA=127 deg, each strip is 0.''46 wide and the fits are performed on the average of the SE and NW axes}
\end{deluxetable}